\documentclass{egpubl}
 
\JournalSubmission    
\usepackage[T1]{fontenc}
\usepackage{dfadobe}  

\usepackage{cite}  
\BibtexOrBiblatex

\electronicVersion
\PrintedOrElectronic
\usepackage{graphicx}
\usepackage{egweblnk} 

\usepackage[T1]{fontenc}
\usepackage{dfadobe}  
\usepackage{booktabs}  
\usepackage{multirow}
\usepackage{amsmath} 
\usepackage{hhline}
\usepackage{mathptmx} 
\usepackage{tabu}  
\usepackage{subfigure}
\usepackage{diagbox}
\usepackage{colortbl}
\usepackage{setspace}
\usepackage{cite}  
\usepackage{titlesec}
\usepackage[normalem]{ulem}

\titlespacing*{\subsection}{0pt}{5pt}{0pt}

\newcommand{\toolName}[1]{\textit{DiffLens}}
\newcommand{\revise}[1]{\textcolor{black}{#1}}

\title{\toolName{}: A Visualization System to Explore Local Differences \\in Graph Sampling}

\author[Zhou et al.]
{\parbox{\textwidth}{\centering Zhiguang Zhou\thanks{Corresponding author}$^{1}$,Yong Zhang$^{1}$, Yuming Ma$^{1}$, Yuqi Zhou$^{1}$, Ke Lu$^{1}$,
Yong Wang\thanks{Corresponding author}$^{2}$, 
Yuhua Liu$^{1}$, \\Jingfang Mao$^{3}$, Yongheng Wang$^{4}$, Ying Zhao$^{5}$, Wei Chen$^{6}$
        }
        \\
{\parbox{\textwidth}{\centering $^1$Hangzhou Dianzi University, Hangzhou, China  \{zhgzhou, 221330016, mamingming, 241330039, luke19999, liuyuhua, \}@hdu.edu.cn\\
$^2$Nanyang Technological University, Singapore  yong-wang@ntu.edu.sg\\
$^3$Zhejiang Company of China National Tobacco Corporation, Hangzhou, China mjf402@126.com \\
$^4$Zhejiang Lab, HangZhou, China  wangyh@zhejianglab.org\\
$^5$Central South University, Changsha, China  zhaoying@csu.edu.cn\\
$^6$Zhejiang University, Hangzhou, China  chenvis@zju.edu.cn
       }
}
}

\begin{document}


\maketitle
\begin{abstract}
Graph sampling techniques have been widely used to simplify network computation and visualization, which also results in inevitable differences between the sampled networks and \revise{the} original networks in terms of nodes, edges and structures. Investigating such differences can inform graph sampling technique users of the pros and cons of different techniques and select the appropriate one, and can also help graph sampling developers evaluate their own technique. However, there are still no systematic ways to achieve such a goal. This paper fills this research gap by first proposing systematic and generic quantitative measures to quantify three categories of graph differences (i.e., neighbor-based, path-based, and structure-based). Built upon this, we further propose \toolName{}, a novel visualization system to help graph sampling developers and users intuitively explore local differences at different regions of their interest within a sampled graph, where three new lens-based visual designs are presented to display the neighbor-based, path-based, and structure-based differences respectively. We conducted two case studies and a user study using real-world network datasets to evaluate \toolName{}. The results confirmed its effectiveness and usability  in helping users explore local differences and compare different graph sampling strategies.


\end{abstract}  

\section{Introduction}\label{Introduction}

\begin{figure}[tb]
  \centering 
  \includegraphics[width=\columnwidth]{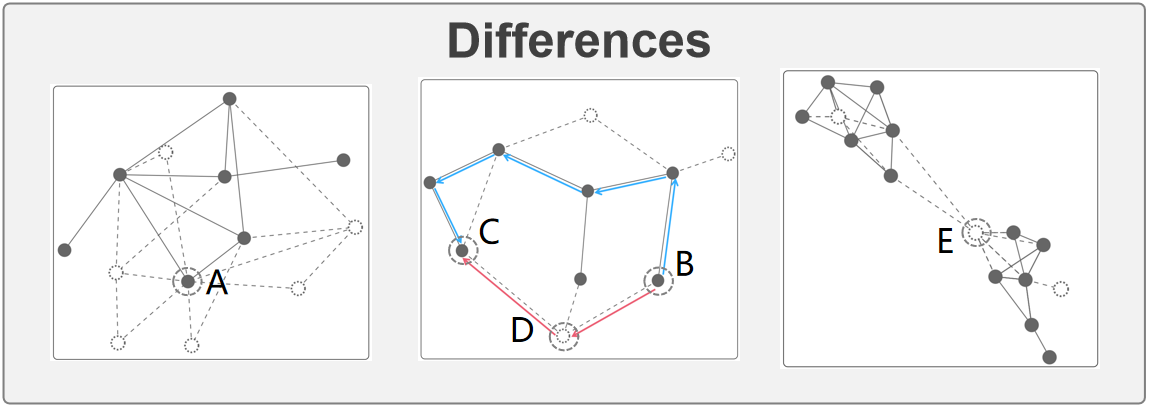}
  \caption{%
Local Structural Changes Induced by Graph Sampling
  }
\label{difference-example}
\vspace{-20pt}
\end{figure}

A variety of graph sampling strategies have been developed to reduce the sizes of large networks, thereby accelerating graph computation and simplifying graph visualization\cite{Guided}. Despite their specific goals and desired properties, it is inevitable to filter out a large number of nodes and edges, thus bringing differences between the sampled and original graphs, such as the changes in node degrees, shortest paths, and community structures. For example, the absence of certain nodes can substantially mislead the estimation of distance, cohesion or other structural measures\cite{article2222}. Fig. \ref{difference-example} illustrates three local regions of a sampled graph. Node A is prominent in the original network, but its influence largely declined in the sampled graph due to the removal of its neighbors. Nodes B and C are a pair of nodes with a short distance, but their distance is enlarged in the sampled graph due to the removal of key bridge node D. When node E is filtered out in the sampled graph, two connected clusters are broken up making more nodes cannot be linked with each other. Such differences are ubiquitous in sampled graphs, which bring considerable uncertainties and errors to network exploration and analysis \cite{articleMissing}.

\begin{figure*}[tb]
  \centering 
  \includegraphics[width=0.85\textwidth]{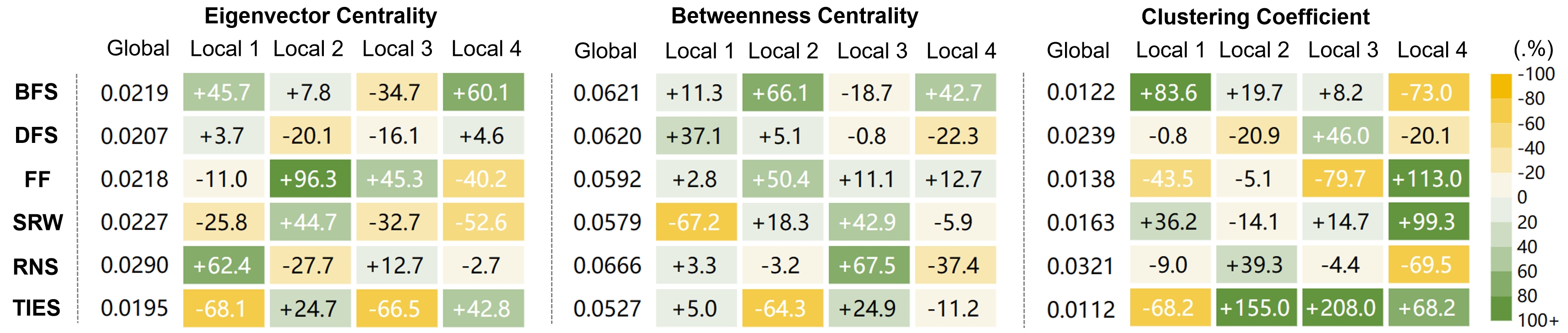}
  \caption{%
  	We performed six sampling strategies (\underline{B}readth-\underline{F}irst, \underline{D}epth-\underline{F}irst, \underline{F}orest \underline{F}ire, \underline{S}imple \underline{R}andom \underline{W}alk, \underline{R}andom \underline{N}odes, \underline{T}otally-\underline{i}nduced \underline{E}dge) on a paper citation data~\cite{repository} and 
 evaluated the sampled graphs based on three classic metrics: Eigenvector Centrality~\cite{article29}, Betweenness Centrality~\cite{dolev2010routing}, and Clustering Coefficient~\cite{article56}. The metrics were calculated and recorded in the entire graph and specific local regions (Local1, Local2, Local3, Local4) within the sampled graph. Then, we calculated the proportion of changes in local differences relative to global differences. All values in local regions are presented in percentage units.}
  \label{chayi}
  \vspace{-20pt}
\end{figure*}

Hence, when performing graph sampling, it's important and beneficial to inform users of the differences between the sampled and original graphs\cite{2013survey}. In fact, plenty of metrics have been proposed to measure the differences in various attributes between sampled and original graphs. 
Zhang et al. ~\cite{zhang2017visual} evaluated graph sampling strategies from different perspectives based on nearly 30 measures (e.g., degree distribution and clustering coefficient). Apart from statistical metrics, certain studies ~\cite{nguyen2017proxy}~\cite{wu2016evaluation} have introduced visual perception criteria for network evaluation, like cluster shape and cluster spatial coverage. Most of these approaches take a global perspective, focusing on the changes in averaged properties. However, in the actual process of graph analysis, users frequently concentrate on the local regions of a graph. For example, in social networks \cite{social}, people are more interested in their friends, relatives, and direct contacts. In urban traffic analysis \cite{traffic}, it is important to capture the traffic dynamics in specific regions. Since network nodes interact with each other more locally, global metrics cannot capture the local differences between sampled and original graphs. Fig. \ref{chayi} illustrates the performance variations among different local regions under the metrics. It can be seen that such global measures fail to accurately depict the performance of the sampled graph. 

Existing global evaluation metrics measure the average performance across the entire graph, often obscuring critical structural changes in specific local regions. As a result, users may overlook poorly-sampled but important subgraphs resulted from sampling strategies. Moreover, global difference scores do not reliably indicate the performance of downstream tasks, such as classification or link prediction, creating a disconnect between graph sampling quality evaluation and their practical utility in downstream tasks. This gap is particularly relevant for sampling developers and users. Sampling developers need to identify where their algorithms introduce local distortions or fail to preserve key substructures, in order to refine or compare different methods. Sampling users, such as data analysts working on social networks or traffic networks, often need to choose strategies that maintain task-relevant local structures—for example, preserving influencer neighborhoods or ensuring routing continuity.  These users require tools that provide interpretable insights into localized sampling differences, helping them detect problematic regions and make informed decisions.   Rather than relying solely on global metrics, such tools should offer intuitive visual feedback on the extent, distribution, and causes of local discrepancies. Lens ~\cite{tominski2014survey} is an efficient strategy to help users conduct detailed examinations on the specific regions of interest, which has been widely applied in a variety of fields, such as geographical analysis~\cite{tominski2012stacking} and medical analysis ~\cite{gasteiger2011flowlens}. It would be a suitable tool for the localized evaluation of sampled graphs, allowing users to gain deeper insights into graph sampling differences. 
Currently, there is little research leveraging lenses for exploring the differences in graph sampling. To fill this gap and implement lenses tailored for graph sampling differences, three challenges will be encountered: \textbf{C1}. There are a large number of graph sampling evaluation metrics. It is complicated and impossible to develop a specific lens for each metric. Therefore, it is essential to derive representative factors that abstract the informational and computational traits of existing metrics, enabling a comprehensive evaluation framework for graph sampling and its integration into the lens for local analysis.
\textbf{C2}. Differences caused by graph sampling are diverse and can vary significantly across local areas. How to define suitable visual mappings for the design of localized lenses is crucial for users to visually capture the differences between the original and sampled graphs. 
\textbf{C3}. We also need to provide a rich set of interactions for the lens, to assist users in exploring different categories of differences and uncovering their underlying causes.

To overcome the above challenges, we begin by conducting a thorough investigation of existing graph evaluation metrics and classify them into three categories (i.e., neighbor-based, path-based, and structure-based).
 Then, we conducted statistical analysis to select the most representative factors for different categories. Further, a set of quantification strategies is designed to measure each category of graph sampling differences. On this basis, we present a novel visualization system named \toolName{} for fine-grained exploration of graph differences between the sampled and original graphs within user-specified areas of interest. It comprises difference-based lenses and, in conjunction with multicolor heatmaps, conducts a multi-level guided exploration. The multicolor heatmaps highlight areas with significant differences, followed by a radial design based on multiple lenses to capture changes in specific factors within sampled graphs. These lenses, offering rich interactions, aid users in identifying and investigating the reasons behind observed differences in graphs. Users requiring graph sampling can leverage our system for visual insights, facilitating the comparison of sampling strategies and the attainment of high-quality sampled graphs. To the best of our knowledge, our work is the first attempt to enhance the localized perception of graph sampling differences. Two case studies and a user study based on real-world networks are conducted to demonstrate the effectiveness of \toolName{}, in facilitating network exploration and the evaluation of graph sampling strategies. The major contributions of our work are summarized as follows:

\begin{itemize}
    \setlength{\itemsep}{0pt}
	\setlength{\parsep}{0pt}
	\setlength{\parskip}{0pt}
    \item We provide a taxonomy of graph sampling evaluation metrics and extract representative factors based on an objective statistical analysis to stand for different categories of graph sampling differences.
    \item We define a new family of quantitative metrics to capture the changes of representative factors between the sampled and original graphs, which can be further applied for the visual mapping of graph sampling differences.
    \item We design a novel interactive system, \toolName{}, enabling users to get deeper insights into the various kinds of local differences in the sampled graphs.
\end{itemize}

\section{RELATED WORK}\label{Related work}

\subsection{Graph Sampling Techniques}

Graph sampling enhances scalability for graph analysis and visualization. In the past few decades, many graph sampling strategies have been proposed to simplify original large-scale networks, which can be categorized into 4 categories~\cite{zhao2020preserving}~\cite{wu2016evaluation}:

\textbf{Node-based sampling.} Random Node (RN) ~\cite{Community} is the most popular sampling algorithm with the uniformity of node selection. It selects a collection of nodes at random from the original graph to produce the sampled graph. In Random Degree Node (RDN) ~\cite{leskovec2006sampling} and Random PageRank Node (RPN) ~\cite{leskovec2006sampling}, the probability of node selection is respectively proportional to its degree and the weight of PageRank. Thus, RDN and RPN can preserve significant graph structure characteristics in the sampled graphs owing to the selection of those important nodes.

\textbf{Edge-based sampling.} Random Edge (RE)~\cite{leskovec2006sampling} randomly picks a subset of edges from the original graph. As a variation of RE, Random Node Edge (RNE) ~\cite{Reducing} selects a random node in advance, and then chooses its linked edges. Similarly, Random Edge Node (REN) ~\cite{rafiei2005effectively} selects edges randomly, and then retains the edges connected to the nodes. However, the above sampling strategies perform poorly on graph connectivity and integrity. To this end, Ahmed et al. ~\cite{Static} provided a graph induction method based on RE, and picked the links from the sampled nodes, preserving connectivity around high-degree nodes.

\textbf{Traversal-based sampling.} To improve connectivity, many scholars propose random walk based graph sampling strategies, such as Breadth First (BF)~\cite{BFS}, Depth First (DF) ~\cite{doerr2013metric}, Snow-ball (SB) ~\cite{goodman1961snowball}, and Forest Fire (FF) ~\cite{leskovec2005graphs}. 
However, these methods follow deterministic paths leading to local deviations, limiting sampling efficiency and coverage. Random Jump (RJ)~\cite{leskovec2006sampling} was introduced to address this issue by randomly selecting nodes with a certain probability in each iteration step.

\textbf{Feature-based sampling.} 
Recently, scholars have proposed graph sampling strategies to address user-specific requirements for feature preservation. Bhatia et al. ~\cite{bhatia2017efficient} introduced Influence Sampling (IS) to better preserve graph node degree distribution and clustering. Du et al. ~\cite{du2019sgp} proposed a social network sampling method SGP for preserving overall community structure stability. To preserve the semantic association feature, Zhou et al. ~\cite{zhou2020context} proposed a context-aware structure preservation method by means of a graph representation learning model. Zhao et al. ~\cite{zhao2020preserving} introduced mino-centric graph sampling (MCGS) to detect and preserve critical minority structures.

It can be seen that plenty of related works focus on the scale reduction of original networks, while overlooking the accuracy and usability of the sampled graphs. In this paper, we design a visualization method to present the local differences for graph sampling evaluation, enabling users to conduct credible network exploration and compare different graph sampling strategies. 

\subsection{Graph Sampling Evaluation}

A great many metrics have been proposed to evaluate the faithfulness of sampled graphs in retaining graph structural properties.

\textbf{Property-based evaluation.} Many properties are widely used for graph sampling evaluation, such as Degree, Closeness Centrality, Clustering Coefficients, and so on. Degree quantifies the count of neighbors around a node. Thus, Degree ~\cite{zhang2017visual} and Degree Distribution ~\cite{leskovec2006sampling} are able to measure the importance of nodes. Average Neighbor Degree (AND) ~\cite{barrat2004architecture} are utilized to gain insights into local neighborhood connectivity. Unlike degree centrality, Eigenvector Centrality (EC)~\cite{bonacich2007some} considers both the quantity and quality of connections between a node and its neighbors. The probability that a node will be connected to another node with similarity is determined by Degree Correlation (DC)~\cite{newman2003mixing}. Other metrics reflect a node’s position in the overall graph through the path between two nodes and then describe the control of the node over the entire graph. Betweenness Centrality ~\cite{dolev2010routing} believes a node will be significant if it is located on the shortest path between other nodes. Closeness Centrality ~\cite{freeman2002centrality} determines the average distance between a node and other nodes, and nodes with high closeness centrality typically possess faster information dissemination capabilities. Mixing Time ~\cite{mohaisen2012measuring} calculates the random walk length required to reach the stationary Markov chain distribution. Other metrics focus on the structure of nodes and edges. Triangle Count~\cite{article30}, Clustering Coefficient ~\cite{article56}, and Transitivity ~\cite{article36} are utilized to evaluate the graph cohesion and the node aggregation. Modularity ~\cite{newman2006modularity} quantifies the network's division into modules (i.e., groups, clusters or communities), assessing the strength of such divisions within a network.

\revise{These evaluation metrics correspond to specific topological patterns in graphs, many of which are also known to be crucial for downstream graph mining tasks. For example, triangle-based structures are widely used in recommendation and link prediction problems, as triadic closure and triangle motifs provide strong signals for inferring potential connections between nodes ~\cite{shin2020bipartite, rossi2019higher}. Similarly, betweenness centrality has long been recognized as an important indicator for routing and network traffic analysis, since nodes with high betweenness lie on many shortest paths and therefore strongly influence information flow in communication networks ~\cite{dolev2010routing, guan2011routing}. These studies suggest that such metrics can also help assess the potential impact of sampling algorithms on downstream tasks.}

\textbf{Perception-based evaluations.} There are also some researches focusing on the visual perception of samples in the context of visualization. For instance, Wu et al. ~\cite{wu2016evaluation} explored how graph sampling strategies affect the perception factors of node-link visualization, including cluster quality, high-degree nodes, and coverage area. Likewise, Zhang et al. ~\cite{zhang2017visual} defined a set of visual perception standards that cover the ability of clusters to maintain size, shape, and quantity, as well as spatial coverage. Nguyen et al. ~\cite{nguyen2017proxy} created several shape-based visual quality metrics using the “proxy graphs” theory.

It is evident that a large number of metrics are available for graph sampling evaluation, making it impossible to visualize all changes simultaneously. Thus, we provide a small number of representative factors to stand for traditional metrics, which are further employed to quantify the differences between sampled and original graphs.

\subsection{Lenses for Local Insights}

Local exploration allows users to delve into specific data areas, revealing important details and patterns that may be obscured by the overall view.
In visualization, while many methods focus on overall presentations, they often overlook detailed local observations. To address this, interactive lenses have been developed ~\cite{tominski2014survey}, enabling users to zoom into areas of interest and explore multi-scale representations while examining local details thoroughly. Magnifying lenses ~\cite{inproceedingsPrecision}, a prime example of Focus+Context lenses, seamlessly integrate focused areas into the surrounding context but may obscure nearby details. Efforts ~\cite{inproceedingsPolyZoom} ~\cite{articleRoute} attempt to place the lens focus outside the viewport, which subsequently disrupts the contextual coherence. To tackle this issue, distortion-based methods offer a smooth transition between the focus and context view. For example, fisheye views ~\cite{articleFisheye} introduce intentional distortion through varying scale functions. In a table lens ~\cite{rao1994table}, localized distortion of cells provides more space for specific rows or columns within the table. In a node-link diagram ~\cite{inproceedingsEdgeLens}, edges between nodes within the focus area can be removed, and nodes connected to the focal node within the lens are pulled in. Although distortion creates seamless views, nonlinear deformation leads to visual instability. A magic lens ~\cite{bier2023toolglass}, on the other hand, is a specialized lens that alters the visual representation of objects within it instead of magnifying their content, allowing for a focused examination of specific attributes.

Interactive lens design facilitates precise visual analysis and deep data exploration. Tominski et al. ~\cite{inproceedingsFisheyeTree} introduced fisheye tree views and composite lenses to reveal hierarchical structures and local details within large graphs. Hurter et al. ~\cite{articleMoleView} proposed MoleView, a semantic lens for interactive exploration of multivariate relational data, linking edges to distinct data ranges based on spatial and attribute contexts. Karnick et al. ~\cite{articleRoute} integrated detail lenses into map routes for finer geographic focus. Kim et al. ~\cite{articleTopicLens} presented TopicLens for precise document corpus topic modeling within specified interests. Zhang et al. ~\cite{zhang2020clusterlens} introduced ClusterLens, revealing the process of interactive map spatial clustering for pattern recognition.
In this paper, we propose a novel lens design to encode the quantified criteria based on representative factors, which can help users visually perceive the local differences between the sampled and original graphs.
\vspace{-11pt}

\section{REQUIREMENT ANALYSIS AND SYSTEM OVERVIEW}
\subsection{ Requirement Analysis}

To develop a visualization system to present and explore local differences in the sampled graphs, we conducted in-depth discussions with two domain experts, E1 and E2. E1 is a scholar with nearly a decade of experience in the field of data visualization, possessing rich expertise in graph visualization and sampling. E2 is a senior data analyst from a world-class IT company. He has undertaken a set of projects for large network applications in the fields of e-commerce, social networking, smart cities, etc. We have collaborated with both experts for nearly two years on this project. Based on the research on large graph sampling and evaluation, \revise{discussions were conducted once or twice per month.} Finally, experts agreed that \toolName{} is valuable for exploring local sampling differences, providing both quantitative comparison metrics and effective and interactive visual guidance. This dual capability enables precise identification of critical structures, enhancing research efficiency.
Their feedback identified the following four requirements:

\textbf{R1. Identify representative factors from evaluation metrics.} A great many metrics have been proposed for graph sampling evaluation, making it challenging to evaluate the sampled graphs with all the differences in metrics. Both the domain experts deemed it necessary to select a small number of representative metrics standing for traditional metrics, which would simplify graph sampling evaluation as far as possible. 
Therefore, conducting a survey and taxonomy of these metrics to extract representative factors can efficiently explore differences between sampled and original graphs.

 \textbf{R2. Quantify graph sampling differences.} It is crucial to inform users about the degree of difference at various locations in the sampled graph. This enables users to pinpoint areas with significant differences, allowing them to focus more on analyzing these critical regions. At the same time, in order to visualize the differences, we must quantify the representative factors and define visual mappings. We should define a new family of quantitative strategies based on the representative factors, to capture the differences.

\textbf{R3. Illustrate explicit information about local differences visually.} Traditional graph sampling evaluation metrics always evaluate the sampled graphs from a global view. A domain expert claimed that it was quite important to focus on the local evaluation of the sampled graphs, otherwise, much ambiguity will be generated in network exploration. Therefore, we should present quantitative differences across local areas of interest with a variety of visual encodings, enabling users to focus on the significant changes.

\textbf{R4. Provide flexible interactions for difference exploration.} After obtaining local differences, users often want to uncover the underlying details behind these differences. As E1 has noted, when observing a significant difference, we need to identify which node's changes caused the difference and whether this node also affects other nodes. Thus, we should provide a rich set of interactions to focus on the nodes or links of interest and explore their changes. It would be valuable to find the causes of differences and inspire suitable graph sampling strategies to fulfill various user requirements.

\subsection{System Overview}

\begin{figure}[h]
  \centering 
  \includegraphics[width=\columnwidth]{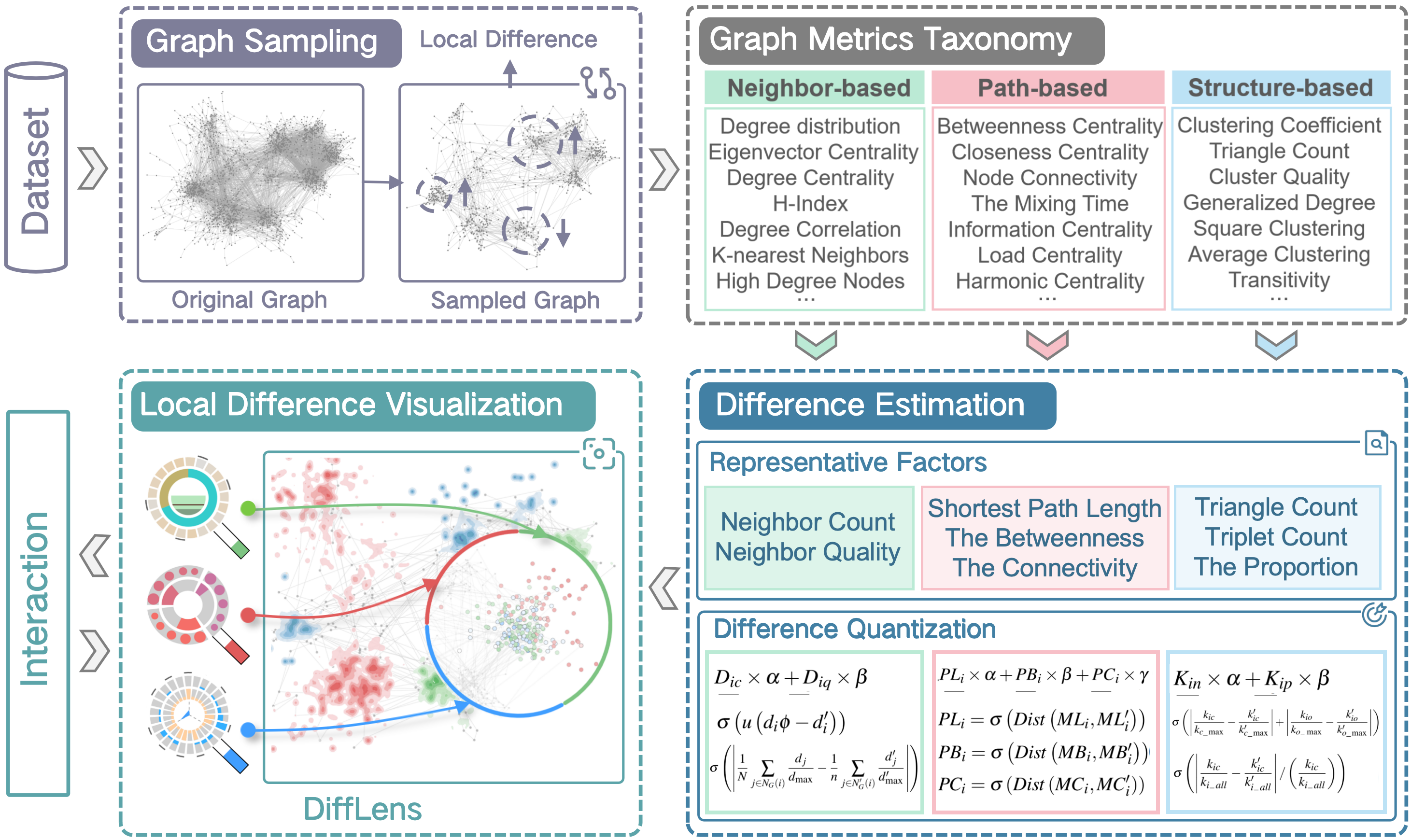}
  \caption{%
  	The pipeline of developing \toolName{}, including the taxonomy of graph evaluation metrics, graph difference estimation, and local visual design of lenses. 
  }
\vspace{-20pt}
\label{pipline}
\end{figure}

\begin{figure*}[tbp]
  \centering
  \mbox{} \hfill

  \includegraphics[width=0.97\linewidth]{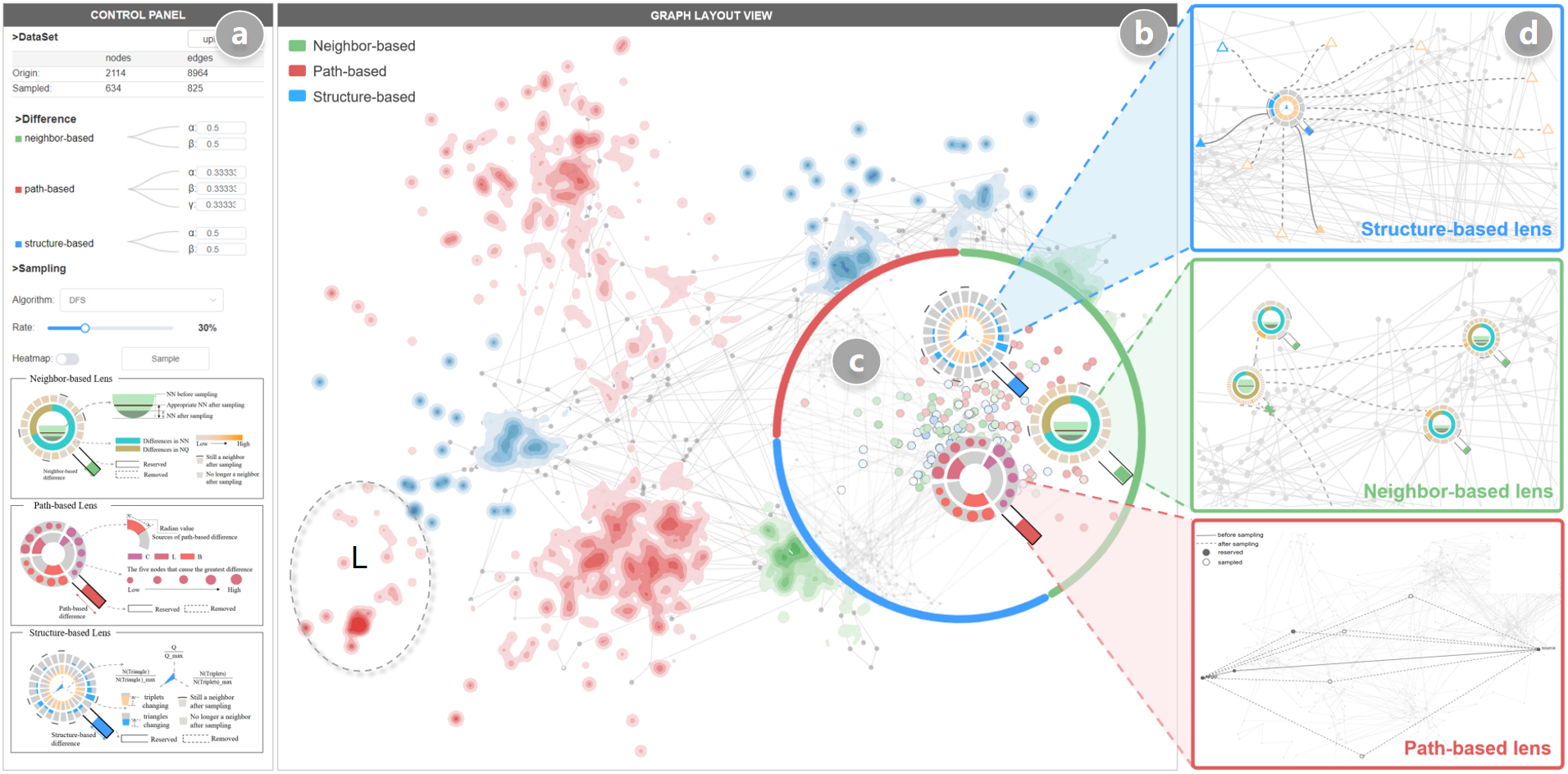}
  
  \hfill \mbox{}
  \caption{\label{system}%
The system interface of \toolName{}: (a) a control panel enabling graph data input and sampling (e.g., strategies, rates, metric weights); (b) a graph layout view in which multicolor heatmaps are overlaid to present the primary differences across local areas; (c) three difference-based lenses presenting three categories of differences within a local area of interest specified in (b); the process of exploring the causes of differences facilitated by lenses, is depicted in (d); (e) categories of differences in three local regions. 
                }
\vspace{-10pt}
\end{figure*}

Motivated by the above requirements, we developed a visualization system, \toolName{}, to facilitate localized difference exploration of sampled graphs. The pipeline of developing \toolName{} is outlined in Fig. \ref{pipline}. First, we conduct a thorough survey on the existing graph evaluation metrics, and categorize local differences into three types: neighbor-based, path-based, and structure-based. Within each category, we summarized the representative factors to capture its characteristics \textbf{(R1)}. Then, we proposed a new family of quantitative methods to estimate local differences based on the representative factors of different categories \textbf{(R2)}. A multicolor heatmap is presented to reveal the categories of differences on the entire graph, and a series of lenses are designed to enhance the visual perception of various local differences in the sampled graphs \textbf{(R3)}. A rich set of interactions are further integrated into the lenses, such as focusing nodes, switching lenses, and moving local areas of interest, enabling users to gain valuable insights into the local differences within the sampled graph \textbf{(R4)}. Fig. \ref{system} presents the visualization interface of our system.

\section{DIFFERENCE CLASSIFICATION}

\subsection{Survey of Graph Sampling Metrics}

We conduct a comprehensive survey on existing graph sampling metrics, including (1) papers published in the journals and conferences in the field of graph visualization and mining, e.g. TVCG, TKDE, CGF, IEEE VIS, EuroVIS, PacificVIS, KDD, JACM; (2) papers published in the IEEE Digital Library identified through a search with the following keywords: ``graph visualization'', ``graph sampling'', and ``graph evaluation''. (3) metrics introduced in the widely-used open-source libraries for graph and network analysis, such as NetworkX. In summary, we collected 63 metrics and observed that the majority of them convey similar meanings, which can be abstracted into eight fundamental factors. 
For example, 19 metrics calculate node degree to present common neighbor characteristics. 24 metrics focus on the pathways between pairs of nodes in the graphs. The other metrics focus on the structural features of the graphs. By collectively considering the literature review on graph sampling evaluation metrics and the insightful feedback from the experts in the expert interview (Section 3.1), we classify these metrics into three categories: neighbor-based, path-based, and structure-based, which are detailed in Appendix A.

\begin{table*}
\centering
\caption{Correlation analysis between our selected factors and conventional sampling metrics.}
\fontsize{9}{13}\selectfont
\label{table1}
\resizebox{\linewidth}{!}{
\begin{tabular}{c|c|ccccccc} 
\toprule
{\multirow{3}{*}{Neighbor-based}} & \diagbox{Factors}{Metrics}      & \begin{tabular}[c]{@{}c@{}}Degree \\Centrality~\cite{article29}\end{tabular}             & \begin{tabular}[c]{@{}c@{}}Eigenvector\\~Centrality~\cite{article29}\end{tabular} &\begin{tabular}[c]{@{}c@{}} H-Index\\~\cite{article1220}    \end{tabular}                                                        & \begin{tabular}[c]{@{}c@{}}Average~Neighbor \\Degree~\cite{barrat2004architecture}\end{tabular} & \begin{tabular}[c]{@{}c@{}}VoteRank \\~\cite{article1844}     \end{tabular}                                               & \begin{tabular}[c]{@{}c@{}}PageRank   \\  ~\cite{zhang2017visual}         \end{tabular}                                           & \begin{tabular}[c]{@{}c@{}}Effective\\ Size~\cite{article15}     \end{tabular}                                              \\ 
\cline{2-9}
{}&Neighbor count      & \textbf{1}                                                              & \textbf{0.8350}                                                  & \textbf{0.8276}                                              & 0.0289                                                            & \textbf{0.6376}                                             & \textbf{0.8991}                                               & \textbf{0.9435}                                                    \\
{}&Neighbor quality     & 0.0288                                                                  & \textbf{0.7892}                                                  & 0.1292                                                       & \textbf{1}                                                        & -0.2389                                                     & -0.1945                                                       & -0.0782                                                            \\ 
\hline
{\multirow{4}{*}{Path-based}} &\diagbox{Factors}{Metrics}         & \begin{tabular}[c]{@{}c@{}}Betweenness \\Centrality~\cite{dolev2010routing}\end{tabular}        & \begin{tabular}[c]{@{}c@{}}Closeness \\Centrality~\cite{freeman2002centrality}\end{tabular}   & \begin{tabular}[c]{@{}c@{}}Node \\Connectivity~\cite{article1215}\end{tabular}  & \begin{tabular}[c]{@{}c@{}}Information\\Centrality~\cite{article1213}\end{tabular}   & \begin{tabular}[c]{@{}c@{}}Local\\Centrality~\cite{article1207}\end{tabular} & \begin{tabular}[c]{@{}c@{}}Reaching \\Centrality~\cite{article1218}\end{tabular} & \begin{tabular}[c]{@{}c@{}}Second~Order \\Centrality~\cite{article29}\end{tabular}  \\ 
\cline{2-9}
{}&Shortest path length & -0.4550                                                                 & \textbf{-0.9783}                                                 & \textbf{-0.7401}                                             & \textbf{-0.8003}                                                  & -0.4582                                                     & \textbf{-0.9666}                                              & \textbf{0.8015}                                                    \\
{}&Connectivity         & 0.4206                                                                  & \textbf{0.7526}                                                  & \textbf{1}                                                   & \textbf{0.9602}                                                   & 0.4252                                                      & \textbf{0.7868}                                               & \textbf{-0.9280}                                                   \\
{}&Betweenness          & \textbf{1}                                                              & \textbf{0.5399}                                                           & 0.4206                                                       & 0.4269                                                            & \textbf{0.9993}                                             & \textbf{0.5447}                                                        & -0.3869                                                            \\ 
\hline
{\multirow{4}{*}{Structure-based}} &\diagbox{Factors}{Metrics}     & \begin{tabular}[c]{@{}c@{}}Clustering \\Coefficient~\cite{article56}\end{tabular} & \begin{tabular}[c]{@{}c@{}}Square\\Clustering\cite{2002The}\end{tabular}      & \begin{tabular}[c]{@{}c@{}}Subgraph\\Centrality~\cite{article1207}\end{tabular} & \begin{tabular}[c]{@{}c@{}}Node Clique \\Number~\cite{article1837}\end{tabular}      & \begin{tabular}[c]{@{}c@{}}Number \\of Cliques~\cite{article1837}\end{tabular} & \begin{tabular}[c]{@{}c@{}}Generalized\\Degree~\cite{article1218}\end{tabular}   &                                                                    \\ 
\cline{2-9}
{}&Triangle count      & -0.0594                                                                 & 0.0677                                                           & \textbf{0.8649}                                              & \textbf{0.7703}                                                   & \textbf{0.8258}                                             & \textbf{0.6937}                                               &                                                                    \\
{}&Triplet count       & -0.2069                                                                 & -0.0534                                                          & \textbf{0.8226}                                              &  \textbf{0.5653}                                                           & \textbf{0.8539}                                             & 0.2935                                                        &                                                                    \\
{}&Proportion           & \textbf{0.9426}                                                         & \textbf{0.5903}                                                  & 0.4168                                                       & \textbf{0.7600}                                                   & 0.4124                                                      & 0.2568                                                        &                                                                    \\
\bottomrule

\end{tabular}}
\vspace{-10pt}
\end{table*}

\subsection{Selection of Representative Factors}
Given the complexity and redundancy among existing evaluation metrics, it would be impractical to visualize or analyze all of them simultaneously. Thus, it is crucial to use representative factors to delineate massive graph  sampling evaluation metrics. We aim to abstract and refine these metrics into a minimal yet expressive set of representative factors to facilitate intuitive interpretation of local graph differences. These factors are not directly chosen from existing metrics, but carefully designed through metric group analysis, simplification, and normalization, providing a clearer conceptual foundation for difference quantification.We first identify three kinds of factors to  represent the basic characteristics of corresponding categories. Their representativeness is confirmed through correlational experiments.

\subsubsection{Neighbor-based Factors}

Neighbor-based metrics are highly dependent on node degree analysis, such as Degree Centrality (DC), Degree Distribution (DD), and Eigenvector Centrality (EC). Therefore, we use neighbor count and quality as the representative factors. Neighbor count is the number of neighbors for a node, while neighbor quality is the ratio of a neighbor's degree to the highest-degree node in the graph.

\subsubsection{Path-based Factors}

Path-based metrics are related to the pathways between pairs of nodes. The average shortest path is commonly applied to represent graph connections. If the average shortest path of a node is smaller than others, the node would play a significant role in communication within the overall graph. Betweenness measures the frequency of a node that lies on the shortest paths between the other nodes. A node with higher betweenness would be crucial for the overall network. Connectivity represents the minimum count of nodes needed to sever the connection between a pair of nodes. If a pair of nodes share greater connectivity, they are closely related to each other. We take the above three factors to stand for the path-based metrics.

\subsubsection{Structure-based Factors}

\begin{figure}[h]
\centering
\hspace{-0.2cm}\subfigure[]{\includegraphics[width=3.5cm]{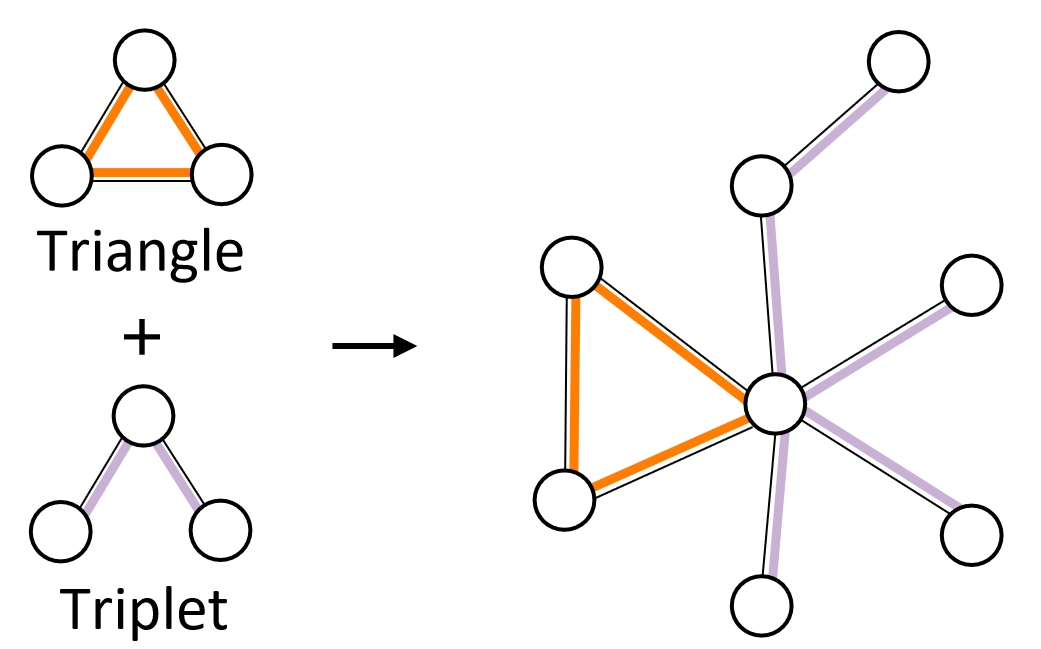}\label{fig:structure1}}
\hspace{0.5cm}\subfigure[]{\includegraphics[width=2.3cm]{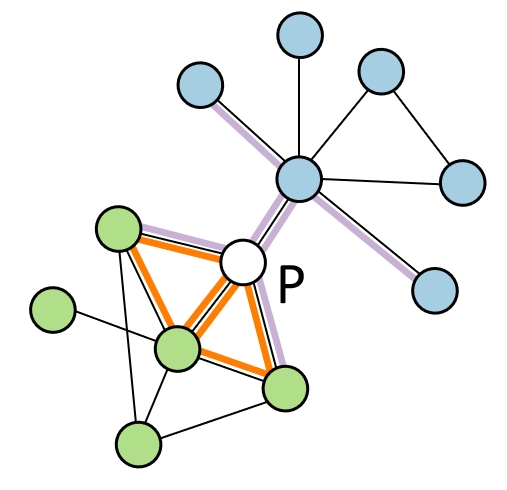}\label{fig:structure2}}

\caption{\label{fig:wholefigure}%
Examples of graph structure exploration based on triangles and triplets. (a) Components of triangles and triplets in graph.(b) Community division assisted by triangles and triplets.}
\vspace{-20pt}
\end{figure}

Many complex structures are contained in the graphs, which are important for specific network exploration applications, such as community and other topological features. On the basis of nodes and connections, we find a close association between triangles (closed triangles) and triplets (open triangles) with these structures. As shown in Fig.  \ref{fig:wholefigure}(a), a structure of interest is made up of two basic units. In fact, all structures in the graphs can be divided into triangles and triplets in the same way. More triangles mean the structures are tightly connected, otherwise, the structures are loosely connected. In Fig. \ref{fig:wholefigure}(b), node P is more likely to belong to the green community because it is the component of two triangles. It can be seen that the distribution of these two units is quite related to the structures of graphs and is important for some specific tasks. Thus, we employ triangle count, triplet count, and their proportion as representative factors of structure-based metrics.


To validate the effectiveness of the representative factors, we conducted a quantitative experiment ~\cite{pearson1895vii}. 
For each category, we selected a subset of metrics for measuring local structural differences before and after sampling based on both common usage and expert recommendations, and computed their Pearson correlation coefficients with representative factors.
Given seven network datasets from Network Repository ~\cite{repository} and SNAP ~\cite{stanford}, the average Pearson correlation coefficients are estimated and recorded in Table \ref{table1}, in which those values bigger than 0.5 are highlighted in bold. It can be seen that each representative factor is highly related to some of the selected metrics in its corresponding category, and all the metrics can be covered based on the representative factors in each category. Therefore, we can take the representative factors to stand for the metrics for further local difference quantification and evaluation.

\vspace{-15pt}
\section{DIFFERENCE QUANTIFICATION}

In this section, we quantify the representative factors to estimate localized differences within the sampled graphs, aiming to cover changes in metrics in the corresponding category through a unified quantitative method. In this paper, the network datasets are all simple, undirected, and unweighted graphs, which are represented as $G=(V, E)$, where $V$ and $E$ are the nodes set and edges set of graph $G$. The sampled graph is denoted as $G'= (V', E')$.

\subsection{Neighbor-based Difference}

Neighbor count and neighbor quantity are selected as the representative factors to stand for the neighbor-based graph sampling evaluation metrics. We define a quantitative method to achieve neighbor-based differences with the selected representative factors. It is worth noting that the representative factors are not computed based on raw differences alone. Their formulations are carefully designed to account for changes in graph scale, the normalization of metric values, and the structural dilution effects introduced by the sampling. This design principle also applies to the path-based and structure-based categories, as will be discussed in the following sections. As shown in Equation (1), the neighbor-based difference is comprised of two components. $D_{ic}$ quantifies the changes of neighbor $c$ount, and $D_{iq}$ quantifies the changes of average neighbor quality. The weights $\alpha$ and $\beta$, initially set to 1/2, can be specified by users to tune the relative impacts of neighbor count and neighbor quality on the derived differences.
\begin{equation}
\mathit{
{Difference}(i)=D_{ic} \times \alpha+D_{i q} \times \beta.}
\end{equation}
$D_{ic}$ is used to quantify the change of neighbor count for a node $i$, which is formulated as:
\begin{equation}\label{gongshi}
D_{i c}=\sigma\left(u\left(d_{i} \phi-d_{i}^{\prime}\right)\right),
\end{equation}
where $d_{i}$ and $d_{i}^\prime$ are the degrees of node $i$ in the original and sampled graphs. $\phi$ is the sampling rate. $\sigma$ is defined to regulate the obtained values into a range between (0,1). $u(t)$ denotes the non-negative portion of $t$, defined as $u(t) = max(0, t)$. $D_{iq}$ quantifies the change of average neighbor quality for a node $i$, which is defined as:
\begin{equation}
D_{i q}=\sigma\left(\left|\frac{1}{N} \sum_{j \in N_{G}(i)} \frac{d_{j}}{d_{\max }}-\frac{1}{n} \sum_{j \in N_{G}^{\prime}(i)} \frac{d_{j}^{\prime}}{d_{\max }^{\prime}}\right|\right),
\end{equation}
where $d_{max}$ and $d_{max}^\prime$ are the maximum node degrees in the original and sampled graphs. $N$ and $n$ denote the \revise{number} of neighbors for node $i$ before and after sampling. $N_{G(i)}$ and $N'_{G(i)}$ represent the sets of neighbors of node $i$ in the original and sampled graphs.

\subsection{Path-based Difference}

Based on the representative factors selected to stand for the path-based evaluation metrics, we define a path-based difference estimation method in Equation (4), where  ${PL}_{i}$, ${PB}_{i}$, ${PC}_{i}$ respectively present the changes of shortest path $l$ength, $b$etweenness, and $c$onnectivity for a node $i$.  $\alpha$, $\beta$, and $\gamma$ are the component weights, initially set to $1/3$, and adjustable based on user requirements. 
\begin{equation}
\mathit{Difference(i)={PL}_i \times \alpha+{PB}_i \times \beta +{PC}_i \times \gamma}.
\end{equation}
How to calculate ${PL}_{i}$,${PB}_{i}$,${PC}_{i}$ is illustrated in Equation (5):
\begin{equation}\mathit{
\begin{array}{l}
\vspace{1.2ex}
P L_{i}=\sigma\left( Dist\left({ML}_{i}, {ML}_{i}^{\prime}\right)\right), \\
\vspace{1.2ex}
P B_{i}=\sigma\left( Dist\left({MB}_{i},{MB}_{i}^{\prime}\right)\right), \\
\vspace{1.2ex}
P C_{i}=\sigma\left(Dist\left({MC}_{i}, {MC}_{i}^{\prime}\right)\right),
\end{array}}
\end{equation}
where $Dist()$ represents the Euclidean distance between two matrices. ${ML}_i, {MB}_i$, and ${MC}_i$ respectively present the matrices of the shortest path, betweenness, and connectivity between node $i$ and the other nodes, which are defined as shown in Equation (6). After graph sampling, they are labeled as ${ML}_{i}^{\prime}, {MB}_{i}^{\prime}$, and ${MC}_{i}^{\prime}$.
\begin{equation}
\begin{array}{l}
\vspace{1.2ex}
ML_{i} = (L_{i1} ,L_{i2} ,...,L_{ij},...,L_{im})^{T}, \\
\vspace{1.2ex}
MB_{i} = (B_{i1} ,B_{i2} ,...,B_{ij},...,B_{im})^{T}, \\
\vspace{1.2ex}
MC_{i} = (C_{i1} ,C_{i2} ,...,C_{ij},...,C_{im})^{T}. 
\end{array}
\end{equation}
\begin{equation}
L_{ij}=\frac{l_{ij}}{l_{\max }},  \quad B_{ij}=\frac{b_{ij}}{b_{\max }}, \quad C_{ij}=c_{ij}.
\end{equation}
$m$ is the count of nodes in $V$, and $l_{ij}$ represents the shortest path length between node $i$ and $j$. $l_{max}$ is the longest length among all the shortest paths in $G$. $b_{ij}$ represents the proportion of shortest paths passing through node $i$ between $j$ and other nodes. $b_{max}$ is the maximum proportion among all pairs of nodes. $c_{ij}$ refers to the minimum count of nodes needed to disconnect node $i$ and node $j$. 

\subsection{Structure-based Difference}

As mentioned above, triangle count, triplet count, and their proportion are representative factors to stand for structure-based evaluation metrics. We define a structure-based difference estimation method based on these representative factors as follows:
\begin{equation}
\mathit{
Difference (i) = K_{in} \times \alpha+K_{i p} \times \beta \\},
\end{equation}
where $K_{in}$ is the changes of triangle count and triplet count formed by node $i$ and its neighbors, which is defined as Equation (9). $K_{ip}$ estimates the change in the proportion of triangles within the two structures, as shown in Equation (10). $\alpha$ and $\beta$ are the weights for the estimation of structure-based difference, which are both initialized as $1/2$ and can be tuned by users.
\begin{equation}
\mathit{
K_{{i} n} = \sigma\left(\left|\frac{k_{i c}}{k_{c \_\max }}-\frac{k_{i c}^{\prime}}{k_{c \_\max}^{\prime}}\right|+\left|\frac{k_{i o}}{k_{o_{-} \max }}-\frac{k_{i o}^{\prime}}{k_{o \_\max}^{\prime}}\right|\right)\\},
\end{equation}
where $k_{ic}$ and $k_{io}$ represent the count of triangles (closed triangles) and triplets (open triangles) formed by node $i$ and its neighbors in $G$. In $G^{\prime}$, they are denoted as $k_{ic}^{\prime}$ and $k_{io}^{\prime}$. ${k_{c_{-}\max}}$, ${k_{o_{-}\max}}$, ${k_{c_{-}\max}^{\prime}}$ and ${k_{o_{-}\max}^{\prime}}$ are respectively referred as the maximum values. 
\begin{equation}
\mathit{
K_{i p} = \sigma\left(\left|\frac{k_{i c}}{k_{i_{-} a l l}}-\frac{k_{i c}^{\prime}}{k_{i_{-} a l l}^{\prime}}\right| /\left(\frac{k_{i c}}{k_{i_{-}  a l l}}\right) \right)\\},
\end{equation}
 where $k_{i_{-}{all}}$ and $k_{ic_{-}{all}}^{\prime}$ are the sum of the numbers of the two structures formed by node $i$ and its first-order neighbors in graph $G$ and $G^{\prime}$. Specifically, when $k_{i c}$ is 0, $K_{i p}$ is 0.

Based on the above quantification processes, the computational complexity of \toolName{} primarily arises from the difference quantification module, which covers three types of sampling differences: neighbor-based, path-based, and structure-based. Among them, path-based differences are the most computationally intensive one, involving the computation of shortest path ($O(|V| + |E|)$), betweenness ($O(|V||E|)$), and connectivity ($O(|V|^2)$) in our practical local-region setting. Overall, the computational complexity is $O(|V||E|)$ for sparse graphs, and $O(|V|^2)$ for dense graphs.

\begin{figure}[tb]
\centering
\subfigure[]{\includegraphics[width=8.2cm]{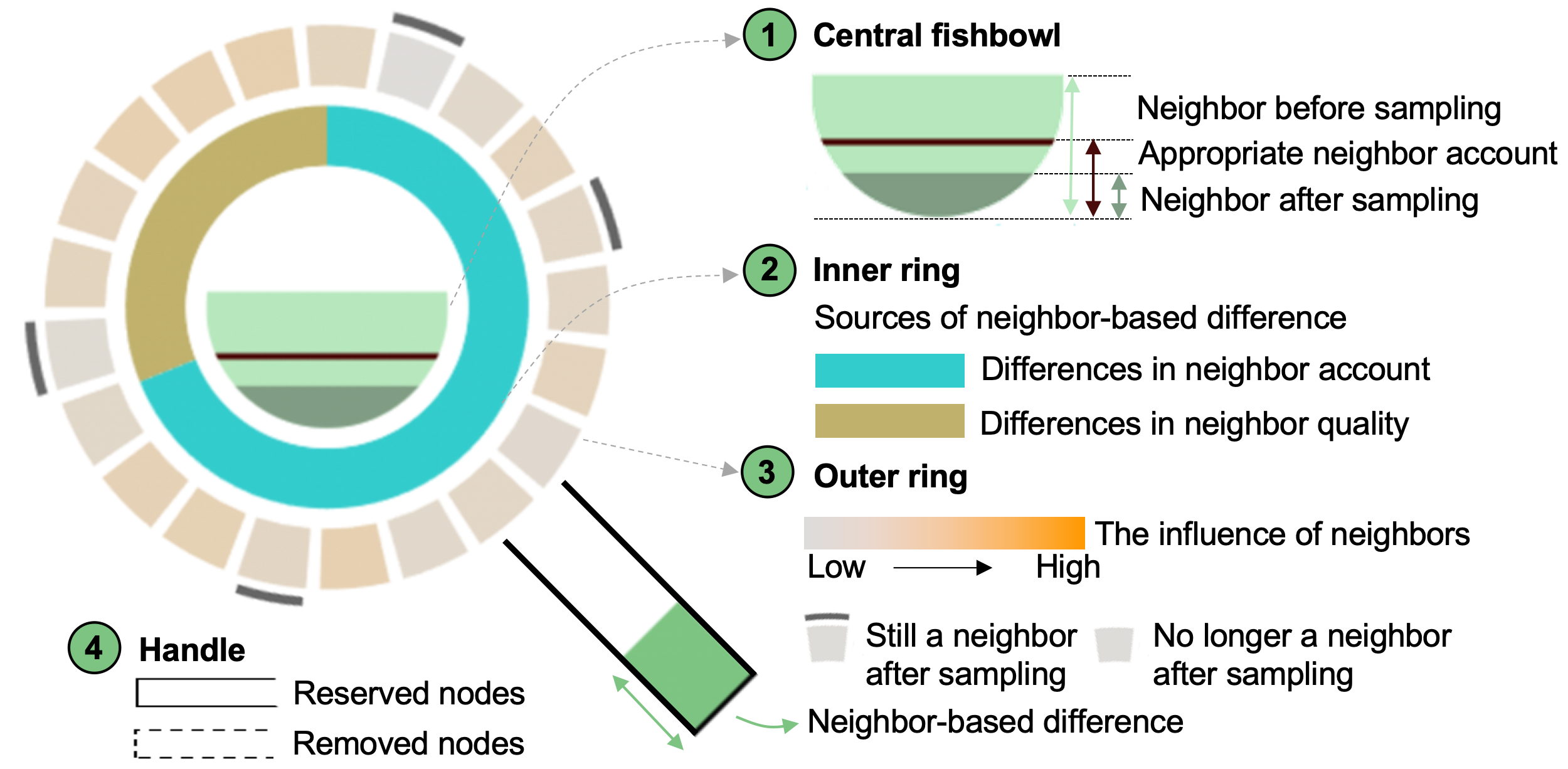}\label{fig:structure1}}
\subfigure[]{\includegraphics[width=8.2cm]{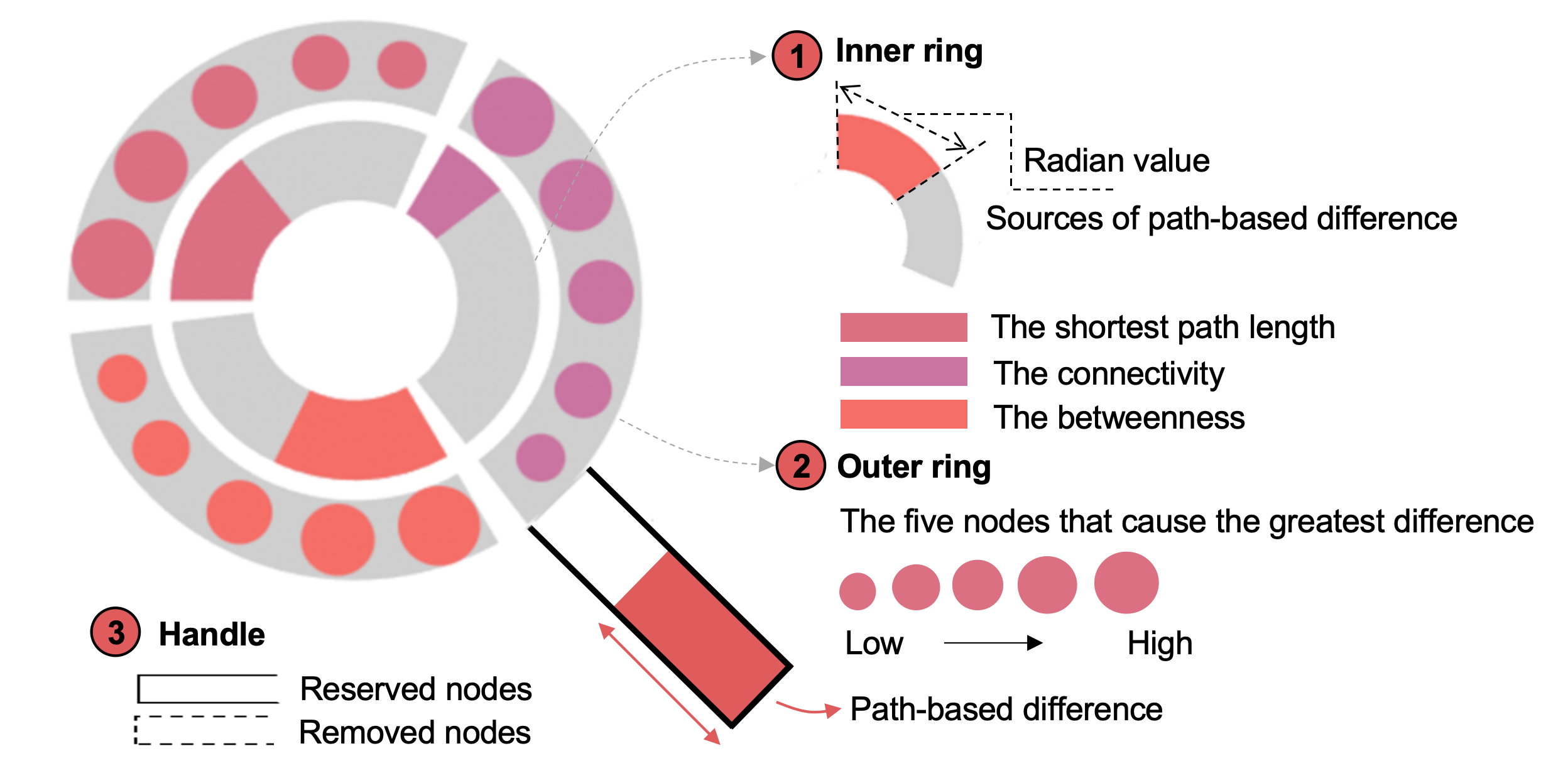}\label{fig:structure2}}
\subfigure[]{\includegraphics[width=8.2cm]{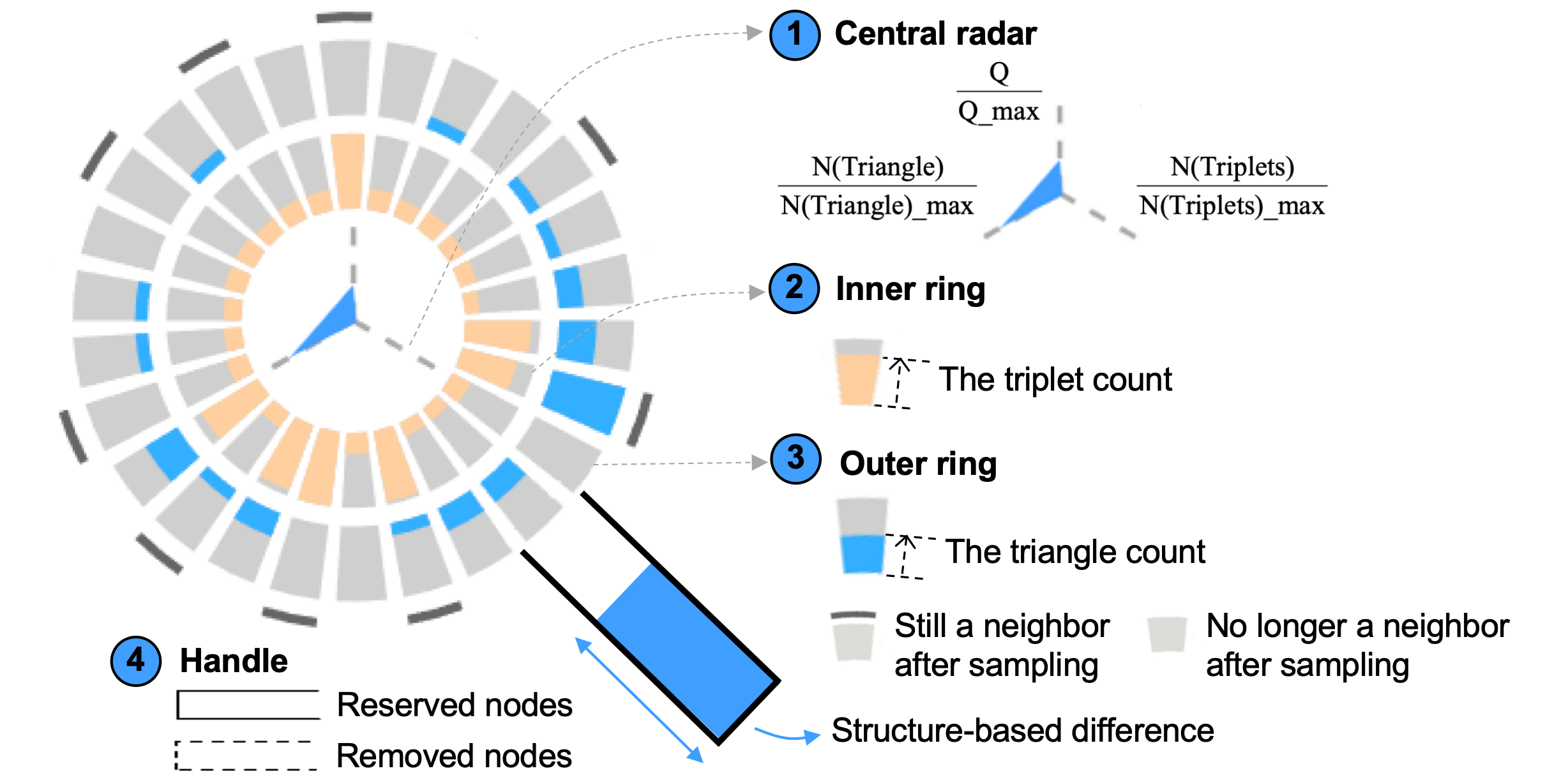}\label{fig:structure3}}
\caption{\label{threelens}%
The design of difference-based lenses. (a) Neighbor-based lens. (b) Path-based lens. (c) Structure-based lens.}
\vspace{-10pt}
\end{figure}



\begin{figure}[tb]
  \centering 
  \includegraphics[width=\linewidth]{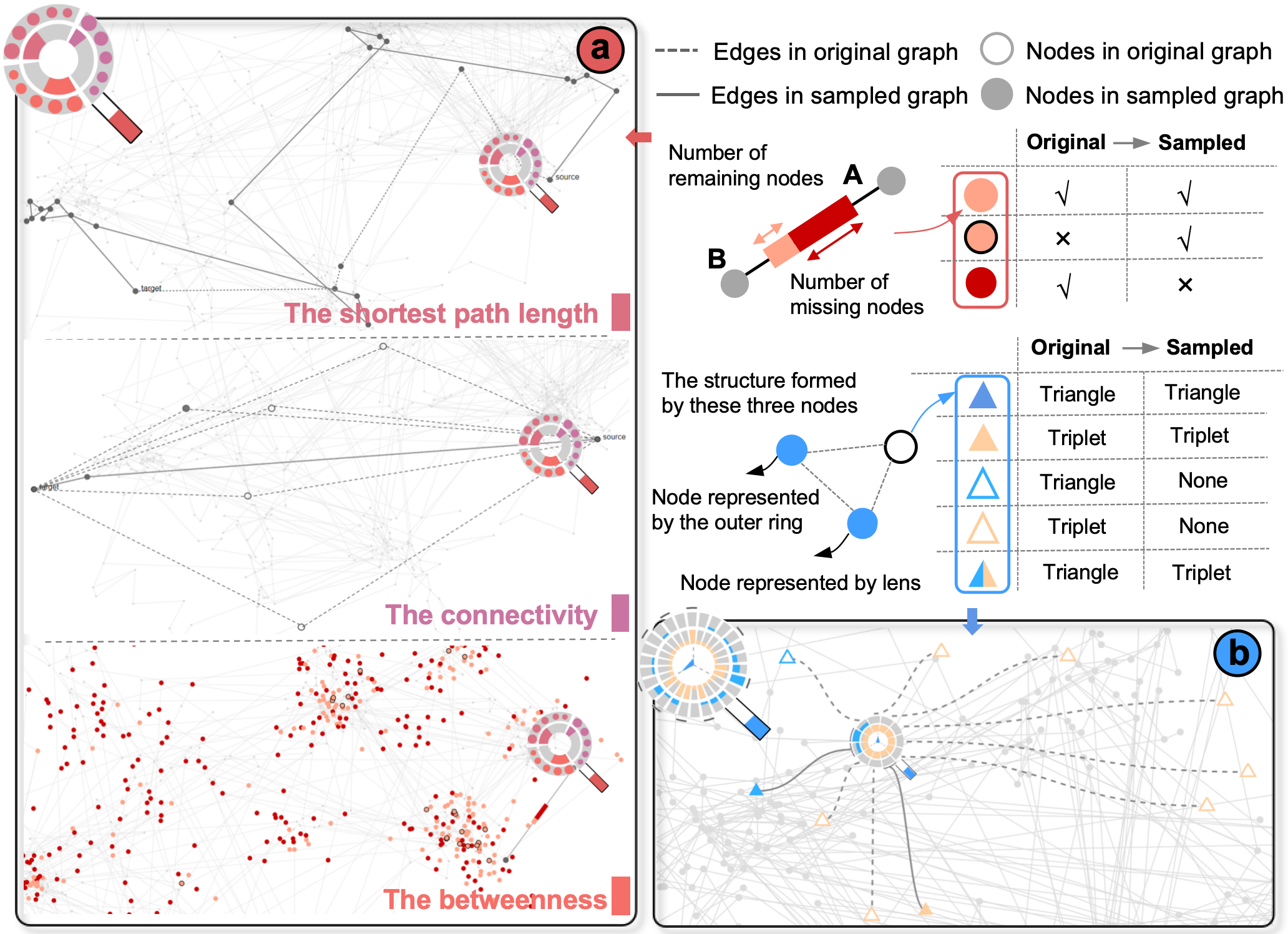}
  \caption{%
  	Interactive designs for local difference exploration based on the path-based lens (a) and structure-based lens (b). 
  }
  \label{pathstru}
  \vspace{-17pt}
\end{figure}

\vspace{-10pt}
\section{DIFFERENCE VISUALIZATION}

\subsection{Graph Layout}

A graph layout view is presented to visualize the network datasets with a node-link diagram. Sampling strategies and sampling rates can be specified by users in the control panel (Fig. \ref{system}(a)) to obtain and perceive their desired sampled graphs in the graph layout view (Fig. \ref{system}(b)). 
For each category of difference, we represented a community's ranking by averaging the global rankings of all its nodes for that difference. The highest-ranked difference was designated as the community’s dominant sampling-induced difference, which was then visually encoded. Given that heatmap\cite{heatmap} is a powerful technique to display the spatial distribution of a specific attribute, we present a multicolor heatmap on the graph layout view to highlight different kinds of graph sampling differences across local areas. \revise{Given that the human visual system has higher perceptual discriminability for the three primary colors,} i.e., red, green and blue \cite{stockman2000spectral}, we employ an RGB color scheme to represent the three differences, thereby enhancing visual discriminability as depicted in Fig. \ref{system}(b).
Community L is shaded in red, which means path-based differences dominate this local area. Similarly, communities dominated by neighbor-based and structure-based differences are shaded in green and blue respectively. Owing to multicolor heatmaps, users can find the global distribution of differences and be guided to focus on the local areas with different kinds of differences.

\subsection{Difference-based Lenses}
We have designed three types of lenses: neighbor-based, path-based, and structure-based, using a unified hierarchical structure and customized visual element filling. Such a design choice takes into account the characteristics of different sampling differences and helps reduce the visual perception load.

\begin{figure*}[tb]
  \centering 
  \includegraphics[width=0.92\linewidth]{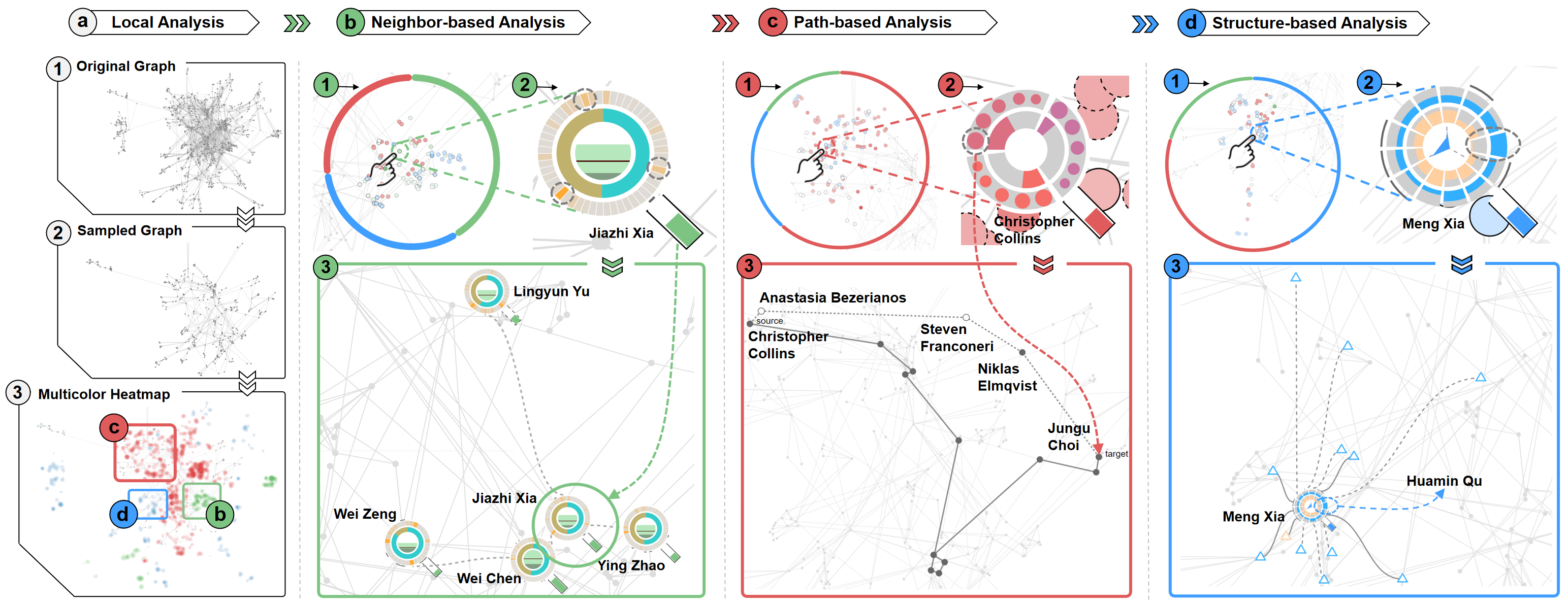}
  \caption{%
  	Heatmaps and lenses for the visualization of local differences in the sampled graphs. (a) The original and sampled graphs are presented, and a difference-based heatmap is laid on the sampled graph. Local areas are shaded in different colors to indicate their dominant differences: neighbor-based differences are shaded in green, while path-based and structure-based differences are respectively shaded in red and blue. (b) (c) and (d) present the interactive exploration of the causes of graph sampling differences based on difference-based lenses.
  }
  \label{case1}
   \vspace{-10pt}
\end{figure*}

\subsubsection{Neighbor-based lens}

We design a neighbor-based lens to encode the changes in representative factors like neighbor count and quality. As shown in Fig. \ref{threelens}(a), the neighbor-based lens is comprised of four components: central ``fishbowl'', inner ring, outer ring, and ``handle''. In the central ``fishbowl'', the heights of sections in the light green and dark green present the neighbor counts for a node of interest respectively in the original and sampled graphs. The ``waterline'' shaded in black is designed to present the expected neighbor count in the sampled graph (original neighbor count × sampling rate), as depicted in equation \eqref{gongshi}. If the section of the dark green falls below the ``waterline'', it means that neighbors are sampled out excessively. 
In the inner ring, two sectors filled with blue and brown respectively represent the differences in neighbor count and neighbor quality, revealing the major causes contributing to neighbor-based differences. In the outer ring, some sectors represent the neighbors of the node of interest, of which the darkness indicates the proportion of a neighbor contributing to the neighbor-based differences. If there is a black line displayed above a sector, it means the neighborhood between it and the node of interest is still retained in the sampled graph. Besides, the length of the green rectangle in the “handle” encodes the neighbor-based differences for the node of interest. A dashed stroke indicates the node of interest has been sampled out. \textbf{Interactions.} In the out ring, when a sector is clicked by users, its neighbor-based lens will generate and a line will be drawn to connect the neighbor-based lenses (a dashed line links disconnected neighbors, while a solid line links neighbors that remain connected after sampling). Users can examine multiple instantiated neighbor-based lenses.

\subsubsection{Path-based lens}

A path-based lens is designed to encode the changes of the representative factors: shortest path length, connectivity, and betweenness. As shown in Fig. \ref{threelens}(b), the path-based lens is comprised of three components: inner ring, outer ring and “handle”. Three components in the inner ring present the proportion of different factors (shaded in the different degrees of red color) contributing to the path-based difference. In the outer ring, five circles in each sector present the most important nodes, and the sizes encode their contributions to the path-based differences. The length of the red rectangle in the “handle” represents the path-based differences for the node of interest. \textbf{Interactions.} In the out ring, when a circle of interest is specified by users, the path-based differences will be generated on the graph layout. The changes in shortest path, betweenness, and connectivity are illustrated through the respective visual designs, as depicted in Fig. \ref{pathstru}(a), enabling users to gain in-depth insights into the causes of different local path-based differences.

\subsubsection{Structure-based lens}

We provide a structure-based lens to encode the changes of representative factors: triangle count, triplet count and their proportion. As depicted in Fig. \ref{threelens}(c), the structure-based lens is comprised of four components: central “radar”, inner ring, outer ring and “handle”. In the central “radar”, three areas respectively present the contributions of difference based on the representative factors, which help users easily identify the major causes of the structure-based differences. In the inner ring and outer ring, \textit{n} sectors are both included corresponding to the neighbor nodes. The heights of orange sectors in the inner ring represent the count changes of triplets based on the corresponding neighbors. Similarly, the heights of blue sectors in the outer ring represent the count changes of triangles based on the corresponding neighbors. The length of the blue rectangle in the “handle” represents the structure-based differences for the node of interest. \textbf{Interactions.} When a sector of interest is clicked in the out ring, the structure-based differences generated from the node and its neighbor are presented with a few kinds of lines and glyphs on the graph layout, as shown in Fig. \ref{pathstru}(b).

\section{EVALUATION}
We conducted two case studies and a user study to demonstrate \toolName{}' effectiveness in evaluating and analyzing local differences in graph sampling. For the two case studies, the eight representative factors of the original graphs were pre-computed, and the metrics of the sampled graphs were calculated in real-time during user interactions. Additional quantitative experiments with multiple sampling strategies and datasets are provided in Appendix B, further demonstrating \toolName{}' effectiveness in exploring local differences.
\subsection{Case Study}

{\bfseries Case 1: Local differences exploration of one sampling strategy}

In the first case, we invited E1 as mentioned in Section 3 to conduct network exploration on a collaboration network formed by the authors of IEEE VIS papers from 2018 to 2022. 1445 author nodes and 5821 collaboration links are included. When the network was laid out based on a node-link diagram, E1 obtained a dense collaboration network as shown in Fig. \ref{case1}(a)-1. To simplify the network visualization, he set the sampling strategy to RW and the rate to 25\%, sampling out many nodes and edges. (Fig. \ref{case1}(a)-2), and a difference-based multicolor heatmap was presented upon the graph layout(Fig. \ref{case1}(a)-3). Plenty of local areas were highlighted in red, meaning that the path-based differences dominated the sampled graph.

After obtaining the sampling heatmap, a dense dark green cluster caught E1’s attention. He selected this area, and a local ring was presented with colorful sections indicating the proportion of categories of differences. Subsequently, one of the dark green nodes exhibited a higher color intensity compared to the others, drawing E1’s attention and leading him to click on it. As the node was clicked within the ring, a neighbor-based lens was timely presented in accordance with an author named Jiazhi Xia (Fig. \ref{case1}(b)). He found that a sharp decrease appeared in Xia's neighbor count within the central ``fishbowl''. Also, numerous neighbors were distributed on the outer ring, while only two of them remained in the sampled graph. A number of high-quality collaborators were sampled out, leading to the neighbor quality of Xia varied drastically, such as Wei Chen (118 neighbors), Ying Zhao (55 neighbors), and Lingyun Yu (36 neighbors). E1 found another author node named Wei Zeng presenting plenty of neighbor-based differences,  caused by missing his important neighbor, Wei Chen, in the sampled network. It is indicated that changes in high-quantity neighbors will create substantial neighbor-based differences, leading to much ambiguity in network exploration. 

\begin{figure*}[h]
  \centering 
  \includegraphics[width=0.88\linewidth]{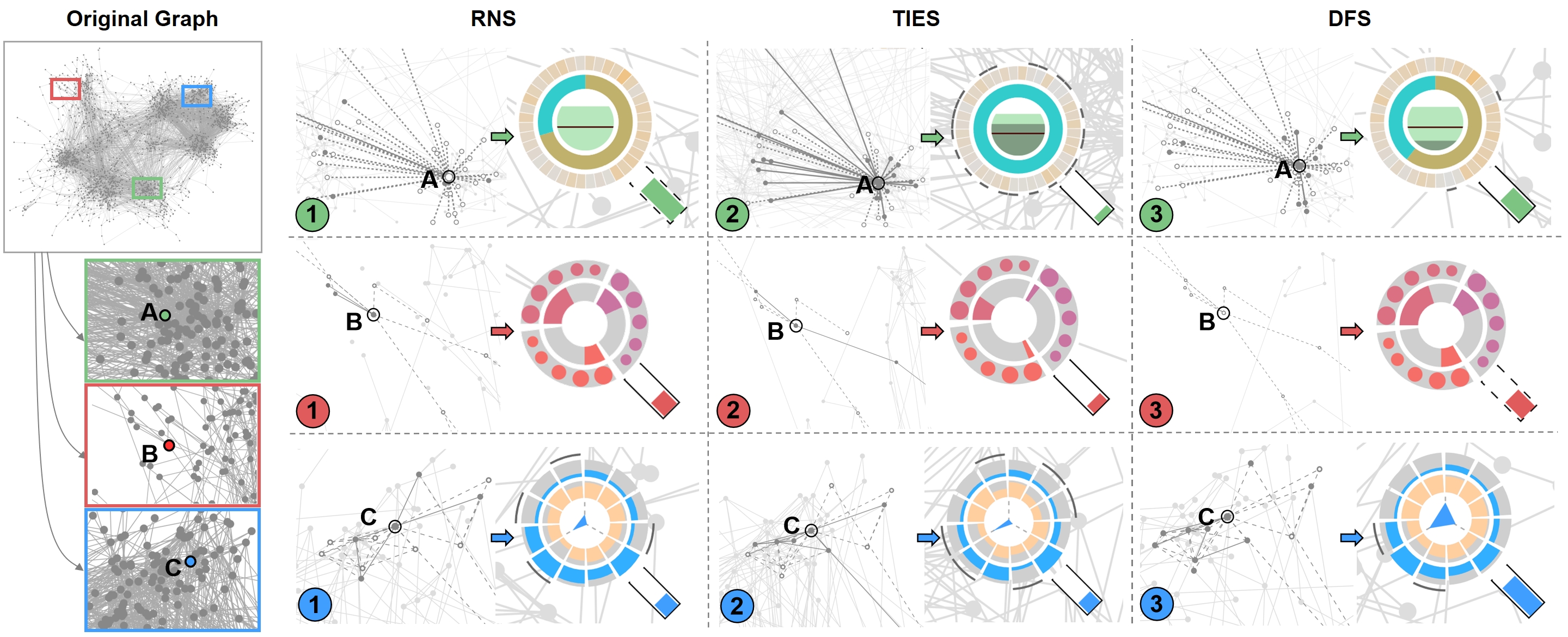}
  \caption{%
   Difference-based lenses focus on local nodes for comparing sampling strategies. Three specific nodes A, B, and C are depicted on the left side. The lens representations of these three nodes under the RNS, TIES, and DFS sampling strategies are shown on the right side.
  }
  \label{lianghua}
   \vspace{-8pt}
\end{figure*}

Then, E1 shifted his attention to a large red-shaded region. Attracted by the color intensity, he selected a path-based lens to focus on a node of interest named Christopher Collins (Fig. \ref{case1}(c)). He found that the main reason for the differences was mostly due to changes in the shortest paths. After clicking on the largest circle named Jungu Choi, he found that the shortest path between them was enlarged greatly because of the absence of Steven Franconeri and Anastasia Bezerianos in the sampled graph. Further, E1 specified a local area shaded in dark blue and selected a structure-based lens  to focus on a node with more differences named Meng Xia (Fig. \ref{case1}(d)).
According to central ``radar'', E1 found the triangle number dominated the structure-based differences of Xia. He specified a section with a larger radius in the outer ring, representing her neighbor named Huamin Qu, and a few blue hollow triangles appeared on the graph layout. It means that Xia and Qu collaborated with many other authors in the original graph, but the triangle relationships disappeared in the sampled graph because of the absence of a link between them.

In this case, E1 aimed to examine how the Random Walk (RW) sampling strategy behaved in preserving key structural elements. During the analysis, E1 observed that RW failed to retain certain important bridge nodes, such as node "Steven Franconeri" and "Anastasia Bezerianos", which disconnected neighboring clusters in the sampled graph. The goal of E1 was to understand these difference patterns in order to identify potential directions for improving the RW strategy. This kind of sampling bias, where transitional nodes or low-degree connectors are underrepresented, motivated E1 to consider how RW could be modified or combined with other techniques to improve its structural fidelity. \toolName{} helped reveal these issues by highlighting differences in local connectivity and triangle density, providing visual cues for understanding specific failure patterns of RW.

After exploration, E1 really appreciated \toolName{} and claimed that ``\toolName{} cannot only help users to visually capture categories of differences, but also guide the in-depth exploration of the causes of differences. With \toolName{}, we can easily trace the origins of sampling-induced differences, which not only builds trust in graph exploration but also provides insights that inspire potential improvements to sampling strategies.''

\noindent{{\bfseries Case 2: Comparison of different sampling strategies} }

In the second case, we invited E2 as mentioned in Section 3 to conduct network simplification based on different sampling strategies and employ our \toolName{} to explore their performance. E2 in this scenario was preparing a sampled subgraph to be used as input for a node classification task. The goal was to select a sampling strategy that retained representative structural features while reducing graph size. Firstly, he loaded a network dataset ~\cite{repository} containing 2.4k nodes and 8.2k edges into the system. A set of widely used graph sampling strategies were selected to reduce the scale of the network, including RNS, TIES and DFS. The sampling rates were all specified as 30\%. 
Given the sampled graphs, E2 began by using a neighbor-based lens to examine the distinctions among various sampling strategies. He focused on a local region dominated by neighbor-based differences and clicked a node A with a notably high degree (Fig. \ref{lianghua}). In the sampled graph generated by RNS, node A was absent, leading to great differences, which is illustrated by the significant green rectangle in the "handle" in the top left corner of Fig. \ref{lianghua}. In addition, most neighbor-based differences stemmed from changes in neighbor quality, as observed in the brown sector. This was due to the loss of high-quality neighbors. E2 argued that such differences were unavoidable because RNS samples nodes with equal likelihood, regardless of their importance. In contrast, node A was retained in the sampled graph based on TIES, and the neighbor-based difference was quite small as shown in the top centre of Fig. \ref{lianghua}. Most of the neighbors were retained in the sampled graph and nearly all the difference originated from the changes of neighbor count. E2 further claimed that ``It meets my expectation that TIES randomly samples edges, preserving high-degree nodes.'' E2 observed that node A remained in the sampled graph generated by DFS. In the top right corner of Fig. \ref{lianghua}, the neighbor-based difference was mainly due to neighbor quality, as most neighbors were not retained except for in-neighbor and out-neighbor through DFS.

Then, E2 specified a path-based lens to focus on node B, which served as a crucial connector linking a subset of nodes with the major parts of the graph. As depicted in the middle left of Fig. \ref{lianghua}, differences induced by RNS were substantial, reflected in the length of the red rectangle in the “handle”, as node B loses connection to the main graph, increasing the shortest path length. In the middle centre of Fig. \ref{lianghua}, E2 observed minimal differences around node B, as a bridging neighbor was retained to connect B to the main graph. Consistent with this, TIES is designed to enhance the connectivity between sampled nodes by adding induced edges. In the middle right of Fig. \ref{lianghua}, node B was excluded through DFS, revealing significant path-based differences resulting from the loss of its neighbors and pathways. DFS utilizes only the local information of nodes, ignoring the area where node B  and its neighbors are located.

Furthermore, E2 specified a structure-based lens to focus on node C in a local area dominated by structure-based differences. In the original graph, it formed a tight structure with 16 triangles and 50 triplets. After RNS sampling, four neighbors are retained, forming one triangle and five triples as shown in the bottom left of Fig. \ref{lianghua}. Accordingly, structure-based differences are presented with the handle and central radar. In the sampled graph of TIES, the handle reveals that the structure-based differences are quite small. And, the central radar indicates that the proportion of two factors is well retained based on four triangles and eleven triples as shown in the middle below of Fig. \ref{lianghua}, indicating robust retention of structures like communities and clustering coefficients. In contrast, DFS sampling retained only one triplet, losing all triangles and significantly disrupting structures, as shown in the lower right corner of Fig. \ref{lianghua}.

Using \toolName{}, E2 compared multiple strategies under the same sampling rate, focusing on how well they preserved neighborhood connectivity, path reachability, and community structure. The comparison revealed that TIES performed consistently well across all aspects, maintaining both hub nodes and structural bridges. Based on these findings, E2 chose TIES to generate a more reliable input graph for training the classification model.

E2 highly affirmed the effectiveness of \toolName{}: “\toolName{} is quite valuable for pinpointing local differences in sampled graphs. We can visually identify differences and explore their causes through encodings and interactions. Owing to such a useful tool, we can handle the differences and timely prompt the potential errors in the network analysis. Also, we can compare  different sampling strategies and select one desired strategy for specific objectives.”

\subsection{User Study}

We further conducted a user study to validate the effectiveness of \toolName{} in exploring the local differences of the sampled graph. 

{\bfseries Participants}. 
We invited 12 graduate students (five females and seven males) with backgrounds in computer science or digital media, and all of them are familiar with visual analysis and graph sampling.

{\bfseries Procedures}. We explained graph sampling categories and factors, introduced three difference-based lenses' designs and interactions, and then demonstrated \toolName{} for exploring local differences. We allowed participants to choose sampling strategies and understand the reasons behind them. 
After the participants were \revise{familiar} with DiffLens, we asked them to use both a simplified version of DiffLens (Baseline) and our full version of DiffLens to conduct graph sampling tasks. For baseline, it enables the selection of sampling strategies and rates, acquisition of sampled graphs, and integration of a rich set of interactions, including zooming, dragging, and color-coding to distinguish sampled nodes.
An email dataset ~\cite{stanford} was utilized in the experiment, including 2114 nodes and 8964 edges. It was sampled based on the sampling rates of 10\%, 20\%, and 30\% by means of TIES, RNS, and DFS. 
Then, the participants  were invited to accomplish the following three tasks respectively with baseline and \toolName{}: \textbf{T1}. Explore the sampled graph from the perspective of node-based factors, and identify five nodes whose removal are considered to have a notable impact and uncertainty on the graph. \textbf{T2}. Explore the sampled graph from the perspective of edge-based factors, and identify five paths whose changes are considered to have a notable impact and uncertainty on the graph. \textbf{T3}. Explore the sampled graph from the perspective of structure-based factors, and identify five structures whose changes are considered to have a notable impact and uncertainty on the graph. Before starting the tasks, each participant was provided with a 10-minute tutorial video that introduced the \toolName{}' main components and demonstrated typical interactions.  After watching the video, participants were given another 10 minutes to freely explore the system and familiarize themselves with the interface.

We recorded the time each user spent on tasks and computed the average values of three classic metrics for the selected nodes, edges, and structures under the two systems, respectively, to compare users’ performance in identifying important sampling changes using the two tools.

\begin{table*}[htbp]
\centering
\caption{The completion performance of user tasks, including the time taken to complete three tasks and the evaluation of results. \\
(DC: Degree Centrality, EC: Eigenvector Centrality, PR: PageRank, BC: Betweenness Centrality, CC1: Closeness Centrality, NC: Node Connectivity, CC2:Clustering Coefficient, NoC: Number of Cliques, SC: Subgraph Centrality.)}

\label{table4}
\fontsize{10}{14}\selectfont
 \resizebox{0.87\linewidth}{!}{
\begin{tabular}{c|c|cccc|cccc|cccc} 
\hline
\multicolumn{2}{c|}{}              & \multicolumn{4}{c|}{10\%}                                                                                  & \multicolumn{4}{c|}{20\%}                                                                                  & \multicolumn{4}{c}{30\%}                                                                                  \\ 
\hline
\multirow{3}{*}{Task 1} &          & {\cellcolor[rgb]{0.937,0.937,0.937}}Time (s)       & DC              & EC              & PR                & {\cellcolor[rgb]{0.937,0.937,0.937}}Time (s)       & DC              & EC              & PR                & {\cellcolor[rgb]{0.937,0.937,0.937}}Time (s)       & DC              & EC              & PR               \\ 
\hhline{~~------------}
                        & BaseLine & {\cellcolor[rgb]{0.937,0.937,0.937}}107.9 & 0.0106          & 0.0823          & 0.0019            & {\cellcolor[rgb]{0.937,0.937,0.937}}119.6          & 0.0093          & 0.0582          & 0.0016            & {\cellcolor[rgb]{0.937,0.937,0.937}}127.9          & 0.0072          & 0.0498          & 0.0014           \\
                        & DiffLens & {\cellcolor[rgb]{0.937,0.937,0.937}}\textbf{106.2}          & \textbf{0.0185} & \textbf{0.1142} & \textbf{0.0044}   & {\cellcolor[rgb]{0.937,0.937,0.937}}\textbf{108.7} & \textbf{0.0157} & \textbf{0.0986} & \textbf{0.0032}   & {\cellcolor[rgb]{0.937,0.937,0.937}}\textbf{116.3} & \textbf{0.0139} & \textbf{0.0911} & \textbf{0.0021}  \\ 
\hline
\multirow{3}{*}{Task 2} &          & {\cellcolor[rgb]{0.937,0.937,0.937}}Time (s)       & BC              & CC1             & NC                & {\cellcolor[rgb]{0.937,0.937,0.937}}Time (s)       & BC              & CC1             & NC                & {\cellcolor[rgb]{0.937,0.937,0.937}}Time (s)       & BC              & CC1             & NC               \\ 
\hhline{~~------------}
                        & BaseLine & {\cellcolor[rgb]{0.937,0.937,0.937}}161.1          & 0.0558          & 0.1799          & 1.9232            & {\cellcolor[rgb]{0.937,0.937,0.937}}167.9          & 0.0396          & 0.1715          & 1.8469            & {\cellcolor[rgb]{0.937,0.937,0.937}}170.1          & 0.0072          & 0.0498          & 0.0014           \\
                        & DiffLens & {\cellcolor[rgb]{0.937,0.937,0.937}}\textbf{122.9} & \textbf{0.0956} & \textbf{0.2135} & \textbf{2.6413}   & {\cellcolor[rgb]{0.937,0.937,0.937}}\textbf{132.3} & \textbf{0.0853} & \textbf{0.1936} & \textbf{2.1564}   & {\cellcolor[rgb]{0.937,0.937,0.937}}\textbf{134.7} & \textbf{0.0139} & \textbf{0.0911} & \textbf{0.0021}  \\ 
\hline
\multirow{3}{*}{Task 3} &          & {\cellcolor[rgb]{0.937,0.937,0.937}}Time (s)       & CC2             & NoC             & SC                & {\cellcolor[rgb]{0.937,0.937,0.937}}Time (s)       & CC2             & NoC             & SC                & {\cellcolor[rgb]{0.937,0.937,0.937}}Time (s)       & CC2             & NoC             & SC               \\ 
\hhline{~~------------}
                        & BaseLine & {\cellcolor[rgb]{0.937,0.937,0.937}}189.1          & 0.2016          & 22.369          & 7846.52           & {\cellcolor[rgb]{0.937,0.937,0.937}}201.9          & 0.1715          & 19.623          & 9843.23           & {\cellcolor[rgb]{0.937,0.937,0.937}}196.4          & 0.0072          & 0.0498          & 0.0014           \\
                        & DiffLens & {\cellcolor[rgb]{0.937,0.937,0.937}}\textbf{134.4} & \textbf{0.5221} & \textbf{31.567} & \textbf{24548.15} & {\cellcolor[rgb]{0.937,0.937,0.937}}\textbf{140.7} & \textbf{0.4469} & \textbf{29.452} & \textbf{22456.56} & {\cellcolor[rgb]{0.937,0.937,0.937}}\textbf{142.5} & \textbf{0.0139} & \textbf{0.0911} & \textbf{0.0021}  \\
\hline
\end{tabular}}
\vspace{-7pt}
\end{table*}


{\bfseries Network-related Task Performance}. We summarized the participants' task completion performance as recorded in Table \ref{table4}. 
It should be noted that these values ~\cite{repository} represent user performance metrics averaged under three sampling strategies.
In addition, the values recorded in the metrics for Tasks 2 and 3 are the averages of the nodes included within the selected paths and structures. \revise{As shown in Table 2, participants consistently completed all tasks faster using \toolName{} than the baseline, indicating that the proposed lenses support more efficient identification and exploration of sampling-induced differences.} Given the statistics, we can observe that the completion time for Task 1 with both tools is similar. It is reasonable that those nodes with larger neighbor-based differences are more intuitively noticeable. For all three tasks, sampled elements identified by \toolName{} showed significantly higher metric values than those from baseline. This demonstrates that users could more accurately locate structural changes of graph sampling using \toolName{}. A paired samples t-test was conducted, revealing a probability ($p$) of 0.036 for equal mean completion times across three sampling rates under two systems. Subsequent tasks yielded $p$ all less than 0.001. After a closer metric examination, it became clear that the participants' selections of nodes, paths, and structures through \toolName{} exhibit more significant differences after sampling compared to the \textit{Baseline}. Similarly, paired samples t-tests, utilizing metrics as features, the results yielded $p < 0.001$, validating \toolName{}' precise identification of nodes responsible for significant differences. \revise{These results suggest that the performance differences between \toolName{} and the baseline are systematic rather than incidental within the controlled study setting.}

\begin{figure}[h]
  \centering 
  \includegraphics[width=\columnwidth]{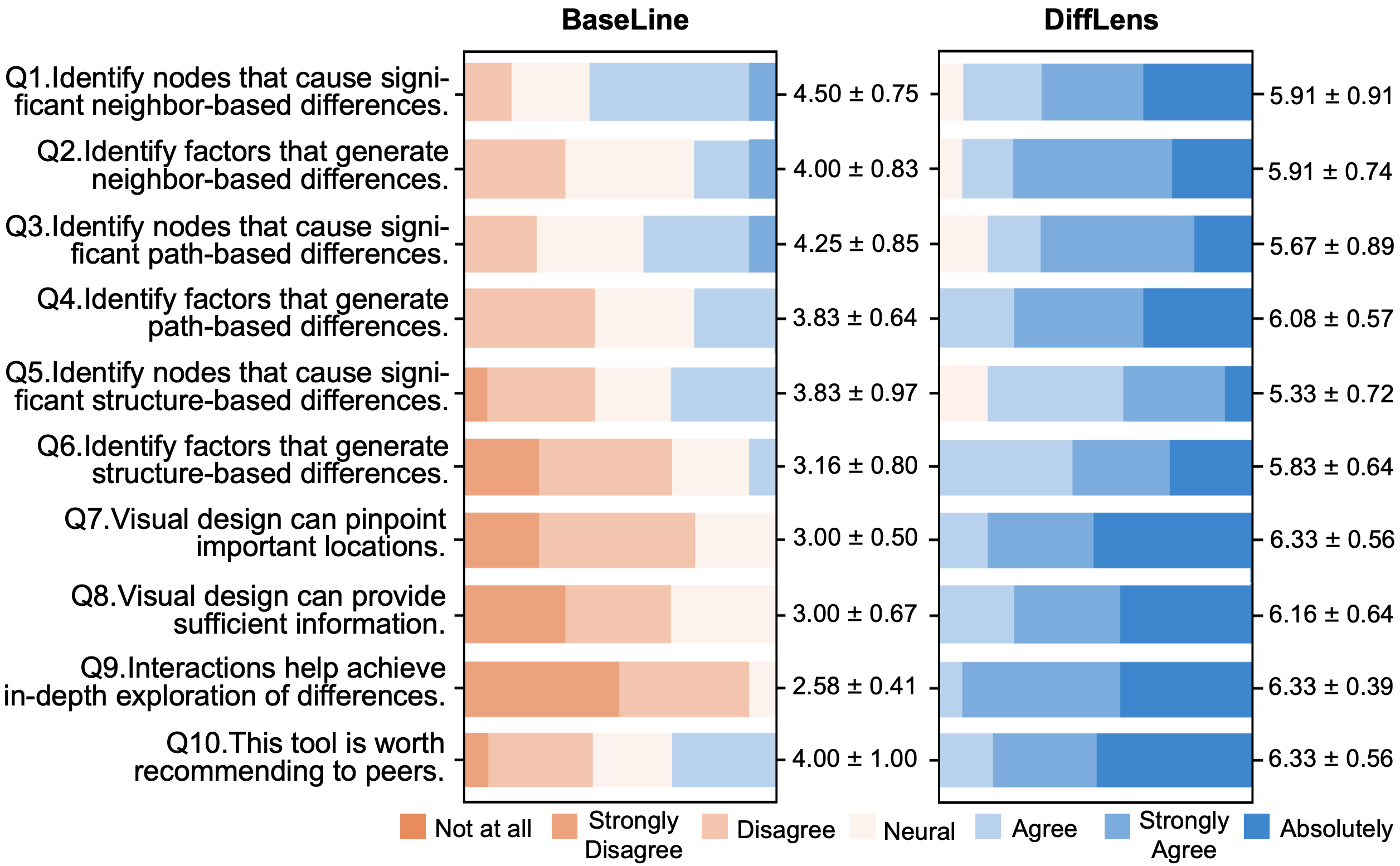}
  \caption{%
The results of Q1-Q10 in our questionnaire under two systems. Our questionnaire focuses on 6 aspects: neighbor-based difference (Q1,Q2), path-based difference (Q3,Q4), structure-based difference (Q5,Q6), visual design (Q7, Q8), interaction (Q9) and system value(Q10).
  }
\label{user}
\vspace{-20pt}
\end{figure}

{\bfseries Post-study Questionnaire Feedback}.
We collected feedback during the tasks, and participants praised the innovative design of our \toolName{}, which enhanced the accuracy of graph assessment with interpretable evaluation criteria. One participant, who spent a large amount of time using the \textit{Baseline}, mentioned that factor selection was the main time-consuming aspect, particularly in Task 3, where identifying the cause of differences was challenging.    Another participant noted that although it took some time to grasp the methods for capturing graph sampling differences and the lens designs, once familiar, the lenses proved to be exceptionally useful. The questionnaire used in the study and participants’ responses can be observed in Fig. \ref{user}. It’s evident that their evaluation of \toolName{} is highly favorable and notably superior to the Baseline, (with yielded $p < 0.001$ in the t-test). \toolName{}, thanks to its ability to offer participants a visual perception of local differences and integrate rich interactions for in-depth insights. As a result, it received ratings exceeding 6 points across Q7-Q9. Furthermore, participants exhibited a strong inclination to recommend our tools to their peers.

\vspace{-8pt}

\section{Discussion and Future Work}

\textbf{Feature-level Difference Estimation.} We estimate differences at the node level using both a multicolor heatmap and \toolName{}.
Features are also significant units, such as community and common topological structures. Thus, the differences in features would be valuable for a variety of network exploration tasks. In future work, we will enhance \toolName{} by designing feature-level difference estimation methods for specific network explorations and applications.

\revise{\textbf{Application-oriented Sampling.} We implement a set of difference-based lenses to show local differences in the sampled graph, thus helping users compare different sampling strategies. However, visualization of differences cannot directly guide graph sampling toward specific application objectives. In future work, it would be interesting to incorporate difference awareness into the sampling process and explore application-oriented sampling strategies, including both improving sampling algorithms and understanding how sampling-induced differences affect downstream tasks.}

\textbf{Semantical Difference Consideration.} The differences identified by \toolName{} are highly related to topological changes in sampled graphs. \revise{In networks where semantic meaning depends on node attributes or edge types, structural differences alone may not fully capture semantic changes introduced by sampling.} \revise{In future work, we plan to extend \toolName{} to incorporate attribute-level and semantic differences, such as node attributes, labels, or domain-specific semantic relations. Such extensions would allow the system to jointly analyze topological and semantic distortions introduced by graph sampling, especially for attributed or heterogeneous graphs.}

\textbf{Scalability.} \toolName{} efficiently handles large node-link diagrams while maintaining manageable computational costs. The experimental results show its capability exploring graph sampling results of graphs with over 5K nodes and 12K edges. However, for spatial graphs where node positions are fixed, visual encoding of \toolName{} may suffer from overlap and limited layout flexibility. Additionally, the current system also displays only one sampling result at a time, which limits side-by-side comparison of multiple strategies. The mitigation of these limitations \revise{is} left as future work.

\vspace{-5pt}

\section{Conclusion}
In this paper, a large number of graph sampling evaluation metrics are investigated and classified into different categories. The representative factors are selected and used to quantify the differences. Then, we propose a novel family of difference-based lenses to encode changes in the representative factors, and present kinds of differences across local areas in the sampled graphs. Given the visual insights and interactions based on the \toolName{}, users would get deeper insights into the local differences and their causes.
 Two case studies and a user study using real-world datasets demonstrate the effectiveness of our system in facilitating network exploration and performance comparison of different sampling strategies.

\section{Ethical Statement}
This research did not involve human participants or animals that would require ethical approval. The user study conducted in this work involved voluntary participation of graduate students in non-sensitive visualization tasks, and therefore ethical approval is not applicable.

\bibliographystyle{eg-alpha-doi} 
\bibliography{egbibsample}       


\newpage

\end{document}